\documentclass[preprint]{aastex}

\newcommand{\Vcom}{V_0}
\newcommand{\Vcirc}{V_{\rm circ}}

\newcommand{\kms}{\,{\rm km}\,{\rm s}^{-1}}

\newcommand{\hubunits}{\,{\rm km}\,{\rm s}^{-1}\,{\rm Mpc}^{-1}}

\newcommand{\Vmax}{V_{\rm max}}
\newcommand{\Vflat}{V_{\rm flat}}
\newcommand{\Mbar}{M_{\rm bar}}

\slugcomment{Submitted to ApJ}
\shorttitle{Dynamical Properties of Disk Galaxies}
\shortauthors{Pizagno et al.}
\begin{document}
\title{Dark Matter and Stellar Mass in the Luminous Regions of Disk Galaxies}
\author{James Pizagno\altaffilmark{1}, Francisco Prada\altaffilmark{2}, 
David H. Weinberg\altaffilmark{1}, Hans-Walter Rix\altaffilmark{3}, 
Daniel Harbeck\altaffilmark{3,4}, Eva K. Grebel\altaffilmark{3,5}, 
Eric F. Bell\altaffilmark{3}, 
Jon Brinkmann\altaffilmark{6},
Jon Holtzman\altaffilmark{7}, 
Andrew West\altaffilmark{8}}
\altaffiltext{1}{Department of Astronomy, Ohio State University, Columbus, OH 
  43210}
\altaffiltext{2}{Ramon y Cajal Fellow, Instituto de Astrofisica de Andalucia 
  (CSIC), E-18008 Granada, Spain}
\altaffiltext{3}{Max-Planck-Institute for Astronomy, Konigst\"uhl 17, D-69117, 
  Heidelberg, Germany}
\altaffiltext{4}{Department of Astronomy, University of Wisconsin, Madison, 
  WI 53706, USA}
\altaffiltext{5}{Astronomical Institute of the University of Basel, 
Department of Physics and Astronomy, 
Venusstrasse 7, CH-4102 Binningen, Switzerland}
\altaffiltext{6}{Apache Point Observatory, 2001 Apache Point Road, P.O. Box 59, 
  Sunspot, NM 88349-0059}
\altaffiltext{7}{Department of Astronomy, New Mexico State University, 
  Box 30001, Department 4500, Las Cruces, N.M., 88003-8001}
\altaffiltext{8}{Department of Astronomy, University of Washington, Box 351580,
  Seattle, W.A. 98195-1580}

\begin{abstract}
We investigate the correlations among stellar mass ($M_*$), disk scale length
($R_d$), and rotation velocity at 2.2 disk scale lengths ($V_{2.2}$) for a 
sample of 81 disk-dominated galaxies (disk/total $\geq 0.9$) selected from the 
Sloan Digital Sky Survey (SDSS).  We measure $V_{2.2}$ from long-slit H$\alpha$
rotation curves and infer $M_*$ from galaxy $i$-band luminosities ($L_i$) and 
$g-r$ colors.  We find logarithmic slopes of $2.60 \pm 0.13$ and $3.05 \pm 0.12$
for the (forward fit) $L_i-V_{2.2}$ and $M_*-V_{2.2}$ relations, somewhat 
shallower than most previous studies, with intrinsic scatter of 0.13 dex and 
0.16 dex, respectively.  Our direct estimates of the total-to-stellar mass 
ratio within $2.2R_d$, assuming a \cite{kro02} IMF, yield a median ratio of 2.4 
for $M_* > 10^{10} M_\odot$ and 4.4 for $M_*=10^9-10^{10}M_\odot$, with large 
scatter at a given $M_*$ and $R_d$.  The typical ratio of the rotation speed 
predicted for the stellar disk alone to the observed rotation speed at $2.2R_d$
is $ \sim 0.65$.

The distribution of scale lengths at fixed $M_*$ is broad, but we find no 
correlation between disk size and the residual from the $M_*-V_{2.2}$ relation, 
implying that the $M_*-V_{2.2}$ relation is an approximately edge-on view of
the disk galaxy fundamental plane.  Independent of the assumed IMF, this
result implies that stellar disks do not, on average, dominate the mass within 
$2.2 R_d$.  We discuss our results in the context of infall models where disks 
form in adiabatically contracted cold dark matter halos.  A model with a 
disk-to-halo mass ratio $m_d=0.05$ provides a reasonable match to the $R_d-M_*$
distribution for spin parameters $\lambda$ ranging from $\sim 0.04 - 0.08$, and
it yields a reasonable match to the mean $M_*-V_{2.2}$ relation.  A model with 
$m_d=0.1$ predicts overly strong correlations between disk size and 
$M_*-V_{2.2}$ residual.  Explaining the wide range of halo-to-disk mass ratios 
within $2.2 R_d$ requires significant scatter in $m_d$ values, with 
systematically lower $m_d$ for galaxies with lower $M_*$
or lower stellar surface density $\Sigma_*$.
\end{abstract}

\section{Introduction}
The tight correlation between luminosity and rotation 
speed is one of the fundamental characteristics 
of the disk galaxy population (\citealt{tul77}, 
hereafter TF).  With stellar population modeling and 
HI gas measurements, this correlation can be expressed 
in terms of stellar mass or total baryonic mass, in 
place of luminosity (\citealt{bel01,mcg00}).
The form and tightness of the TF relation are 
critical tests for theoretical models of galaxy formation 
(e.g., \citealt{col89,kau93,col94,eis96,ste99,avi98,fir00}).
In classic models of 
disk galaxy formation by dissipative gravitational 
collapse (e.g., \citealt{fal80,gun83,dal97,mo98}, hereafter MMW),
the quantities
that determine the disk rotation curve are 
the concentration parameter of the dark matter halo, 
the ratio of the disk baryonic mass to the total mass 
of the halo, and the disk scale length, which is determined 
by its angular momentum.  These theoretical models suggest that 
disk size could be an important additional parameter in 
disk galaxy correlations (\citealt{she02}; 
\citealt{dut05}).  A strong correlation between disk size 
and TF residual is also expected if disks are ``maximal'' 
and therefore make a dominant contribution to the observed 
rotation speed (\citealt{cou99}).  

In this paper, we investigate the correlations among rotation speed, 
stellar mass, and scale length in a sample of disk galaxies selected from the 
Sloan Digital Sky Survey (SDSS; \citealt{yor00,aba04}).  
We have obtained long-slit 
optical spectra of these galaxies and used them to extract H$\alpha$
rotation curves.   We use the $g$-$r$ color of each galaxy to 
estimate its stellar mass-to-light ratio 
$M_{\ast}/L$, and therefore its stellar mass,
 following the prescription 
of \cite{bel03}.  We apply bulge-disk decomposition 
to the SDSS $i$-band images to select a sample of 
disk-dominated galaxies (disk/total $\ge$ 0.9) and 
to measure the disk exponential scale length.  We estimate circular 
velocities at 2.2 disk scale lengths, where the rotation curve 
of a self-gravitating exponential disk 
reaches its maximum (\citealt{fre70}).  
We use the estimated stellar masses to separate the disk and 
halo contributions to the total mass within this radius, 
and we investigate the halo-to-disk ratio as a function of 
stellar mass and disk scale length.  We discuss our results in the 
context of MMW-style disk galaxy models.

\section{Photometric and Spectroscopic Observations}
The SDSS galaxy redshift survey 
has an unprecedented combination of large area, depth, and 
photometric quality, thanks to the combination of a large 
format camera \citep{gun98}, high throughput multi-object spectrographs, 
careful calibration procedures \citep{fuk96,hog01,smi02}, 
and an efficient series 
of data reduction and targeting 
pipelines \citep{lup01,sto02,str02,bla03a,pie03,ive04}.  
We have selected a sample of $\sim$
200 galaxies from the main galaxy sample (\citealt{str02}) 
of the SDSS redshift survey for follow-up dynamical 
study with long-slit H$\alpha$ spectroscopy.  Our full
sample covers a representative selection of galaxies in 
the absolute magnitude range $-18 \ge M_{r} \ge -22$, with 
no morphological pre-selection other than an $i$-band isophotal 
axis ratio cut of $b/a$ $<$ 0.6, needed to allow accurate 
inclination corrections to observed rotation velocities.  We 
impose a minimum redshift, $cz$ $\ge$ 5000 ${\rm km\,s^{-1}}$, 
so that peculiar velocities do not cause large uncertainties 
in distance (and thus luminosity and size). We adopt a 
luminosity-dependent maximum redshift of $9000\kms$
($-18\ge M_{r}> -19.5$), $11000\kms$
($-19.5\ge M_{r} > -20.5$), and $15000\kms$
($-20.5\ge M_{r}$), so that galaxies are spatially resolved 
and the distribution of absolute magnitudes is roughly 
flat over the range $-18$ to $-22$.  Analysis of the Tully-Fisher 
relation and its residuals for this broadly representative 
galaxy sample, including detailed discussion of the sample 
definition, spectroscopic observations and data reduction 
procedures, and rotation curve fits, will be presented by 
Pizagno et al. (in preparation, hereafter P05).
Here we summarize the relevant 
aspects of these procedures and describe the selection of 
the disk-dominated galaxy sample 
that is analyzed in this paper.  

In a series of observing runs between June 2001 and April 2004,
we obtained long-slit spectra covering the H$\alpha$ 
wavelength region for a total of 234 galaxies in the velocity 
and absolute magnitude ranges described above, using the 
TWIN spectrograph on the Calar Alto 3.5-m telescope (189 galaxies) 
and the CCDS spectrograph on the MDM 
2.4-m telescope (45 galaxies).  Typical exposure times 
were 30 minutes at Calar Alto and 60 minutes at MDM, with
instrumental setups yielding FWHM resolution $\sim 1.48$\AA\, and 
$\sim 1.93$\AA\, respectively.
We obtained usable H$\alpha$ rotation curves for 170 
galaxies, $\sim 70\%$ of the input sample, with the remainder showing 
insufficient extended H$\alpha$ emission or 
(in a few cases) excessively irregular velocity profiles.  We
applied the bulge-disk decomposition program 
GALFIT (\citealt{pen02}) to the $i$-band images of these 
170 galaxies, taken from the SDSS corrected frames.  Specifically, 
we fit each galaxy with a combination of an inclined exponential 
disk and a bulge with a surface brightness profile 
${\rm exp} [-(r/r_s)^{1/n}]$ (\citealt{ser68}), with the index $n$ 
constrained to the range 0.5 $\ge$ $n$ $\ge$ 5.0.  

For this
paper, we select those galaxies with disk-to-total luminosity 
ratio $f_d \ge$ 0.9.  These systems may not be perfectly 
described by smooth exponential disks, but the addition 
of a bulge containing more than 10\% of the light does not 
allow a better fit.  We rejected five galaxies for which the 
discrepancy between the GALFIT exponential disk position angle and the 
SDSS isophotal position angle, used for the long-slit 
observations, would lead to a velocity difference of more than 10\%.  
We confirmed the disk-dominated nature of the remaining 81 
galaxies by visual inspection.  
We note that $f_d \ge 0.9$ is a stronger morphological cut
than that in most TF samples, which also include some galaxies
with significant bulges.  We adopt the more stringent cut
mainly because it allows us to define scale length and velocity
measures, $R_d$ and $V_{2.2}$, that are insensitive
to ambiguities of bulge-disk decomposition.  Also, while the
bulge formation mechanisms in late-type galaxies are uncertain,
our sample gives theoretical modelers a clear target to make
predictions for: nearly bulgeless galaxies, with an absolute
magnitude distribution that is approximately flat in the range
$-18 \leq M_r \leq -22$.  We compare results from this sample
to those of our full, morphologically representative sample in P05.

We compute galaxy luminosities using SDSS Petrosian fluxes and 
colors using SDSS model colors, both K-corrected to redshift 
$z=0$  using Blanton et al.'s (\citeyear{bla03b})
{\tt kcorrect\_v3.1b}.  We compute distances using the 
SDSS heliocentric redshifts corrected to the rest frame 
of the Local Group barycenter (\citealt{wil97}),
assuming a cosmological model with $\Omega_m$=0.3, 
$\Omega_{\lambda}$=0.7, and Hubble constant 
$H_0 = 70\hubunits$.
We incorporate a distance uncertainty corresponding to $300 \kms$
when calculating disk scale 
length and luminosity uncertainties, to account 
for the typical amplitude of small scale peculiar 
velocities (\citealt{str95}).  Figure 1 shows the 
distribution of our sample galaxies in the color-magnitude 
plane.  Crosses show galaxies that did not
have enough H$\alpha$ emission for extended 
rotation curves.  Although most of these failed galaxies 
are either red or low luminosity, the galaxies with successful 
rotation curve measurements cover all populated 
areas of the color-magnitude plane, so there are no 
major categories of galaxies in this absolute magnitude 
range that are missing from our sample.  Filled circles show 
the disk-dominated subset analyzed in this paper.  
These again span all populated regions of the color-magnitude 
plane, though the distribution is somewhat bluer than that of
the full sample, and the fraction of disk-dominated systems is higher 
at low luminosity.  The surface brightness distribution (not shown) 
is similar
to that of the full sample, though the low and intermediate luminosity
galaxies that are eliminated by the $f_d \geq 0.9$ cut tend
to be above the median surface brightness.  The GALFIT total 
magnitudes in $i$-band are 0.12 magnitudes brighter
than SDSS Petrosian magnitudes on average, with a standard deviation
of 0.10 magnitudes, in reasonable agreement with expectations
(see \citealt{str02,graham05}).
We use the Petrosian magnitudes for our analysis so that our results
refer to quantities easily accessible from the SDSS database.

The spectroscopic data were dark subtracted, 
flat-fielded, and linearized using standard IRAF 
procedures, as outlined by \cite{mas92}.  
Following \cite{cou97}, we extract 2-D spectra 
along the spatial direction and measure the H$\alpha$ 
emission line centroid at each location along the slit.  The 
H$\alpha$ emission line centroid 
uncertainty is between 2 and 12 $\kms$ depending on the 
signal-to-noise ratio of the emission line.  We define the 
rotation curve as the spatial variation of the emission line centroids 
along the major axis of the galaxy.  
We fit galaxy rotation curves with an arc-tangent 
function, which has a minimal number of parameters while still 
describing the global shape of typical rotation 
curves adequately (\citealt{cou97}).  Specifically, we 
use a Levenberg-Marquardt $\chi^2$ minimization 
routine (\citealt{pre92}) to fit the data with 
the functional form
\begin{equation}
V(r)=\Vcom + \frac{2}{\pi}\Vcirc \mbox{ }{\rm arctan} \left (\frac{r-r_0}{r_t} \right),
\end{equation}
where $\Vcom$ is the central velocity, $\Vcirc$ is the asymptotic 
circular velocity, $r$ is the position along the slit, $r_0$ is the 
center of the rotation curve (where $V=V_0$), 
and $r_t$ is the turnover radius at which the 
rotation curve begins to flatten.  When performing the fit, 
we add 10 ${\rm km\,s^{-1}}$ in quadrature to 
the observational error on each H$\alpha$ data point, to account 
for non-circular motions and to ensure that parameters 
are determined by the overall shape of the 
rotation curve rather than the high signal-to-noise data points in the 
inner parts of the rotation curve with small uncertainties.  

We adopt the circular velocity at 2.2 
disk scale lengths as our measure of rotation speed.  
The rotation curve of an isolated exponential disk peaks at this
radius \citep{fre70}, and \cite{cou97} shows that this velocity 
measure produces the tightest TF relation.
We infer the observed rotation velocity
($V_{2.2}^{\rm obs}$) by evaluating equation (1) at $r=2.2 \times R_d$, where
$R_d$ is the $i$-band disk scale length determined by GALFIT.  
The uncertainty in $V_{2.2}^{\rm obs}$ is determined using the 
covariance of the parameter errors returned by the Levenberg-Marquardt 
routine.  We correct $V_{2.2}^{\rm obs}$ for inclination 
by using the GALFIT-determined disk axis ratio and the equation 
\begin{equation}
V_{2.2} = V_{2.2}^{\rm obs}\left(\frac{1-(b/a)^2}{1-0.19^2}\right)^{-1/2},
\end{equation}
where 0.19 is the assumed intrinsic axis ratio for an edge-on disk and 
$b/a$ is the measured $i$-band axis ratio. Observational 
estimates of the intrinsic axis ratio vary 
from 0.10 to 0.20 depending on galaxy type 
\citep{hay84}.  We choose 0.19, typical for spiral galaxies, and note 
that the range $0.10-0.20$ corresponds to a small variation 
(typically $\sim$1$\kms$) in $V_{2.2}$.  As discussed in 
detail by P05, roughly $1/3$ of our galaxies have rotation 
curves that are still rising at the outermost H$\alpha$ 
point.  We have included these galaxies in our sample, but 
we have checked that excluding them makes minimal difference 
to our results.

Table~1 lists the SDSS identifier, distance, $i$-band luminosity,
$g-r$ color, stellar mass, GALFIT disk fraction, disk exponential 
scale length, and rotation velocity $V_{2.2}$ for the 81 galaxies
that comprise our disk-dominated sample.  
The luminosities and colors are corrected for internal extinction, 
and stellar masses are computed from these extinction-corrected
quantities, as described in the next section.

\section{Results}
Figure 2a show the $i$-band TF relation for our sample of 81 
disk-dominated galaxies.  Following standard practice, 
we correct luminosities, and colors, for internal extinction based on the 
disk axis ratio and luminosity, using the prescription of 
\cite{tul98} interpolated to the central wavelength 
of the SDSS $i$-band.  We convert luminosities to solar units 
using $M_{i,\odot}$ = $4.56$ (\citealt{bel03}). The three 
representative error crosses in the lower right show 
the 90th-percentile, 50th-percentile, and 10th-percentile 
values of the observational uncertainties.  We add $1/3$ of the 
inclination correction in quadrature to the luminosity 
uncertainty to represent the uncertainty in the inclination 
correction itself.  
Luminosity uncertainties are dominated by this inclination correction
uncertainty and by the $300\kms$ peculiar velocity uncertainty.

The solid line shows our ``forward'' fit to the observed 
TF relation.  Specifically, we fit a relation
\begin{equation}
y=a(x-x_0) + b
\end{equation}
with $y={\rm log}\,L_i/L_{\odot}$, $x={\rm log}\,V_{2.2}/{\rm km\,s^{-1}}$, 
assuming a Gaussian intrinsic scatter of dispersion $\sigma$ 
in $y$ at fixed $x$, in addition to observational uncertainties.  
We determine maximum likelihood values of $a$, $b$, and $\sigma$ 
by maximizing
\begin{equation}
{\rm ln}\,(L) = -\frac{1}{2} 
  \sum_i {\rm ln}\, (\sigma^2 + \sigma_{i,y}^2 + a^2\sigma_{i,x}^2) - 
  \sum_i \frac{(ax_i+b-y_i)^2}{2(\sigma^2+\sigma_{i,y}^2+a^2\sigma_{i,x}^2)}
  +\,{\rm constant}~,
\end{equation}
where $\sigma_{i,x}$ and $\sigma_{i,y}$ are the observational 
uncertainties for data point $i$ (see P05 for discussion).  
We determine the $1\sigma$ errors on $a$, $b$, and $\sigma$ by 
repeating this procedure for 100 bootstrap subsamples of the
full data set and taking the dispersion as the uncertainty. 
We choose the reference value $x_0$ so that there is 
essentially no covariance between the errors in 
the slope $a$ and intercept $b$.  

For the forward TF relation, we obtain
\begin{equation}
(L_i/10^{10} L_{\odot}) = {1.84(\pm0.09)}
  \left (\frac{V_{2.2}}{149.6\kms} \right)^{2.60\pm0.13} ~.
\end{equation}
The intrinsic scatter is $\sigma$=0.13 dex = 0.33 magnitudes, 
comparable to that of other TF studies 
(e.g., \citealt{cou97,kan02}).
The dotted line shows the inverse TF fit, in which 
we assume that there is Gaussian intrinsic scatter of ${\rm log}  V_{2.2}$ 
at fixed ${\rm log}  L_{i}$ instead of the reverse.  Forward and inverse 
fits correspond to different hypotheses about the intrinsic 
distribution of the correlated quantities, and they yield 
different slopes except in the limit of zero intrinsic scatter.  
Our inclusion of the intrinsic scatter as a fit parameter means that
points with small {\it observational} errors do not get inappropriately
large weights in determining the slope and normalization, a
difference from many previous analyses.  However, for this sample
the intrinsic scatter is small enough that the derived slope is not
highly sensitive to the fitting procedure (for example, the inverse
fit is $V_{2.2} \propto L_i^{1/2.9}$).
Our slope is shallower than that found by some previous studies,
e.g., \cite{ver01}, who finds a slope of
$\approx 4.5$ in $K'$-band, or \cite{kan02},
who find a slope of $\approx 4.0$ in $R$-band.
However, it agrees well with the $r$-band slope
of $2.54$ found by \cite{cou97}, whose sample criteria and analysis
procedures are most similar to ours.  We discuss possible 
contributions to slope differences in more detail below.  
Our estimates of the slopes, intercepts, and intrinsic scatter 
of these and all other bivariate relations fit in this paper are 
listed in Table 2.  
Residuals from these relations show no discernible correlation with
axis ratio, which indicates that our inclination corrections are 
accurate in the mean, even if they are not perfect on
a galaxy-by-galaxy basis.  The Appendix presents a 
Monte Carlo test for Malmquist-type
biases in our sample selection and analysis and shows that they
are small compared to our quoted statstical errors, with an effect
$\sim 0.01$ on the forward TF slope.

The point types in Figure 2a encode galaxy color, relative to 
the expectation for the galaxy's
$i$-band luminosity.  We fit a linear 
mean relation to the sample's ($g$-$r$) vs.
${\rm log}\,  L_i$ relation and divide the sample 
into three nearly equal parts based on the residual from this 
relation.  Squares represent galaxies redder than the mean 
by 0.02 mag or more, circles represent galaxies bluer by 0.05 mag 
or more, and triangles show the remaining galaxies.  There is 
a slight tendency for red galaxies to lie below the mean TF relation, as 
one might expect given the higher mass-to-light ratios or 
red stellar populations, but the trend is weak relative 
to the scatter.  

In Figure 2b, we have converted galaxy $i$-band 
luminosities to stellar masses, using the prescription 
of \cite{bel03} to infer each galaxy's stellar 
mass-to-light ratio from its ($g$-$r$) color.  
We use color in preference to the spectroscopic methods 
of \cite{kau03} because the SDSS fibers 
cover only the central regions of these relatively 
nearby galaxies and may not sample a representative 
stellar population.  \cite{bel03} 
adopt a ``diet Salpeter'' initial mass function (IMF) 
chosen so that stellar disks have the maximum mass allowed by 
rotation curve constraints.  At each $g$-$r$ color, we 
multiply their stellar mass-to-light ratios by 0.71 to 
correspond to a Kroupa IMF, which better 
represents direct observational estimates of the 
IMF (see \citealt{bel03} for further discussion).  
Specifically, we calculate stellar masses using the relations
\begin{mathletters}
\begin{eqnarray}
M_{\ast}/ M_{\odot} & = & (L_i / L_{\odot})\, (M_{\ast} / L_i) ,\\
{\rm log}\, (M_{\ast}/L_i) & = & -0.222 + 0.864 \times (g-r) + {\rm log}\, 0.71,
\end{eqnarray}
\end{mathletters}
where the two coefficients are from Table 7 of \cite{bel03}.
We use inclination corrected luminosities and colors, again based on 
the \cite{tul98} prescriptions, but the inclination effects are small 
(typically less than 10\%) because
galaxies move along a locus of roughly constant $M_*$ as their luminosities
and colors are corrected for extinction.  We add an error contribution
to $M_*$ that is one-third of the applied inclination correction.

Points in Figure 2b are again coded by galaxy $g$-$r$ 
color residual, now computed as a function of stellar mass. 
The conversion to stellar mass has removed the slight trend of 
TF residual with color residual, as one might expect 
if stellar mass is the more fundamental quantity.  
Solid and dotted lines show our best-fit forward and
inverse relations, which are listed in Table 1. 
The forward relation is
\begin{equation}
(M_{\ast}/10^{10} M_{\odot}) =
  {2.32(\pm0.10)}
  \left(\frac{V_{2.2}}{155.6\kms}\right)^{3.05\pm0.12}.
\label{eqn:mv}
\end{equation}
The best-fit intrinsic scatter is 0.16 dex, slightly higher 
than 0.13 dex found for the $L_i-V_{2.2}$ relation.
Some of this increase could reflect galaxy-to-galaxy variations
in stellar populations or extinction properties, which would change
the true $M_*/L$ ratios at fixed $g-r$ color; we have assumed a
deterministic relation between $g-r$ and $M_*/L$ and have not
included any scatter about this relation in our observational
error budget.  The BD01 models have 0.1 dex scatter in 
stellar mass-to-light ratio ($M_*$/$L_i$) at fixed color.  

The short-dashed line shows the best-fit stellar-mass 
TF relation found by \citeauthor{bel01} (\citeyear{bel01}, hereafter BD01),
using \citeauthor{ver01}'s (\citeyear{ver01}) data for the Ursa Major 
cluster.  We have multiplied the 
normalization of their relation (the $I$-band 
fit with mass-dependent inclination correction 
from their Table 2) by 0.71 to adjust to the Kroupa 
IMF assumed here.  Distances are calibrated to $H_0 \approx 70\hubunits$
in both cases, though there could be some uncertainty in the relative
distance normalization from the peculiar velocity of Ursa Major.
The two relations agree at $V_{2.2} \sim 200\kms$, but the
BD01 relation is substantially steeper than ours, with a slope of 
$4.49\pm0.23$ vs. $3.05\pm0.12$.  
At $V_{2.2} \sim 100\kms$ the BD01 relation traces the lower 
envelope of our data points and lies $\sim 0.4$ dex
below our best-fit relations.  
Differences in the samples and analysis methods include:
our use of the updated \cite{bel03}
stellar population models in place of the BD01 models,
our use of a disk/total $\geq 0.9$ cut vs.\ BD01's more
generic ``late type galaxy'' selection, 
our use of $V_{2.2}$ as a velocity measure in place of
\citeauthor{ver01}'s (\citeyear{ver01}) $\Vflat$
measure used by BD01, and our ``field''
(or, more accurately, random) environment selection vs.\ 
their cluster sample, 

Since \cite{ver01} finds a steep $K'$-band TF relation for the
Ursa Major galaxies, BD01's steep slope (relative to ours) 
appears to be intrinsic to the sample, not a consequence of any
differences in stellar population modeling.
Our disk/total cut also seems unlikely to be the main source of difference,
since the low-$V_{2.2}$ galaxies that {\it are} in our
sample lie significantly above any of the BD01 data points,
and the relatively small number of low-$V_{2.2}$ galaxies excluded by
our cut lie on or above the best-fit relation.
To investigate the importance of velocity definition differences, we used the
data of \cite{cou97}, who lists $V_{2.2}$ and the maximum velocity
$\Vmax$ derived from a 5-parameter fit to optical rotation curves.  
The typical ratio of $\Vmax$ to $V_{2.2}$ is higher for less massive
galaxies, and since $\Vflat$ is likely to track $\Vmax$ more 
closely than $V_{2.2}$, the trend goes in the right direction to
explain the discrepancy.  However, if we scale
up our $V_{2.2}$ values using a mean correction derived from 
the \cite{cou97} data, then our $M_*-V_{2.2}$ slope changes to
3.47, still much shallower than BD01, and the gap
between the relations is still $\sim 0.2$ dex in $M_*$ at
$V_{2.2} \sim 100 \kms$.  Thus, it appears that velocity 
definition differences can account for roughly
half of the difference between our results and BD01's.  We 
reach a similar conclusion by comparing Verheijen's (2001) 
$r$-band TF data points to our own (see P05).  
The most plausible source we can identify for the remaining
gap is a systematic difference in properties of field and cluster
spirals at low luminosity.  Reconciling the two measurements
requires the cluster galaxies to rotate faster by $\sim 0.1-0.2$ dex
at fixed $M_*$.  Fully addressing this difference requires a large 
sample with a range of environments and both HI and optical data, so that one 
can mimic selection and analysis procedures used by different 
authors.   

Our stellar mass TF relation is also shallower than the
{\it baryonic} TF relation derived by \cite{mcg00},
$\Mbar/10^{10} M_\odot = 2.12 (V/155.6\kms)^{3.98}$ 
(for $H_0 = 70\hubunits$).
Here $\Mbar$ is the sum of the stellar mass and the gas mass inferred
from HI measurements.  We cannot directly estimate $\Mbar$ for our 
galaxies because we do not have HI data, but \cite{kan04} reports
a statistical correlation (with substantial scatter) between 
gas-to-stellar mass fraction and SDSS $u-r$ color,
$\log (G/S) = 1.46 - 1.06(u-r)$.  If we apply this correction to
our sample on a galaxy-by-galaxy basis, we obtain
$\Mbar/10^{10} M_\odot = 2.86 (V/155.6\kms)^{2.89}$,
shallower than equation~(\ref{eqn:mv}) because low mass
galaxies have higher gas content.  
The average mass increase is $0.2$ dex at $V_{2.2} \sim 100\kms$
and $0.05$ dex at $V_{2.2} \sim 200\kms$.  
Our slope is substantially shallower than that found by 
\cite{mcg00}, who combine several data sets obtained in 
different bands.\footnote{\cite{mcg00} also assumed a constant
{\it stellar} mass-to-light ratio in each band.  Had they 
included a color dependence, they would have derived a still
steeper slope, since fainter galaxies are generally bluer and 
therefore have lower $M_*/L$.}
The \cite{mcg00} sample covers a much wider mass range than ours,
extending to circular velocities $V \sim 30\kms$.
The difference between optical linewidth $2V_{2.2}$ and the
HI linewidth $W_{20}$ used by \cite{mcg00} could become
more important at these low velocities, partly explaining
the difference in slope, but we do not see an easy way to
fully reconcile the results.

Figure 3 plots the GALFIT disk scale length 
$R_d$ against stellar mass.  The dotted line 
shows the best-fit mean relation
\begin{equation}
R_d = 3.87(\pm0.11) 
  \left(\frac{M_{\ast}}{2.21 \times 10^{10} M_{\odot}}\right)^{0.24\pm0.03} 
  {\rm kpc}~.
\end{equation}
However, scatter about this mean relation is very broad.  
Points are coded by the residual from this best-fit relation, 
with squares, triangles, and circles representing the largest,
intermediate, and smallest $1/3$ of the galaxies at a given 
luminosity.  

Theoretical expectations for the distribution of disk 
galaxies in the space of stellar mass, scale length, and 
circular velocity are clearly described by, e.g., \cite{fal80}, 
MMW, \cite{dal97}, \cite{mo00}, \cite{she02}, and 
\cite{cou03}.  In this paper, we use a modeling approach
similar to that of MMW to place our observational results in
theoretical context.  The central solid line in Figure 3 shows the predicted 
$R_d-M_{\ast}$ relation for galaxies with a ratio 
$m_d$=0.05 of stellar mass to total halo mass formed in an 
NFW halo (\citealt{nav97}) with concentration 
parameter $c$=10 and spin parameter $\lambda$=0.06. We 
compute this relation using equation (28) of MMW, which 
includes the effects of disk self-gravity and adiabatic 
contraction of the inner regions of the halo.  We set the
specific angular momentum of the disk equal to that of 
the halo ($j_d/m_d=1$ in MMW's notation), so our quantity 
$\lambda$ is equivalent to their $\lambda'$.  While 
the predicted relation is steeper than our best fit, it 
roughly describes the central trend of our data points.   
Upper and lower solid lines show the predictions for 
$\lambda$=0.08 and 0.04, respectively.   The envelope 
of these lines encloses roughly the central 80\% of the 
data points.  

The distribution of halo spin parameter in N-body 
simulations is approximately log-normal with a 
mean $\langle \lambda \rangle \approx 0.04$
and dispersion $\sigma_{{\rm ln}\lambda} \approx 0.5$
\citep{bul01}.  For $m_d$=0.05, reproducing 
the $R_d$-$M_{\ast}$ relation requires $\lambda$ values 
in the upper half of this distribution, so disks would either 
have to form in the higher spin halos or have a higher 
specific angular momentum than the dark matter.  Since 
systems with low angular momentum may be more likely 
to form a substantial bulge, and thus be omitted from our 
disk-dominated sample, this preferential sampling of the 
high end of the $\lambda$-distribution is not implausible.  
However, the three solid lines in Figure 3 can also be 
produced (almost exactly) with a disk-to-halo 
mass fraction $m_d$=0.025 and $\lambda$ of 0.03, 0.045, 0.06; 
the lower $m_d$ shifts the predicted relations to 
lower $M_{\ast}$, and lower $\lambda$ values are required 
to compensate.  Conversely, for $m_d$=0.10, the $\lambda$ 
values that yield similar $R_d-M_{\ast}$ curves 
are 0.055, 0.08, 0.11.  
 
Figure 4 again shows the stellar-mass TF relation, 
$M_{\ast}$ vs. $V_{2.2}$, but points are now coded 
by their residual from the best-fit $R_d-M_{\ast}$ 
relation, with squares, triangles, and circles 
representing the largest, intermediate, and smallest 
galaxies, respectively, just as in Figure 3.  There is 
no evident separation among these three sets of points, i.e., 
no tendency of large or small galaxies to lie 
above or below the mean $M_{\ast}-V_{2.2}$ relation.  A 
plot of $M_{\ast}-V_{2.2}$ residual against 
$R_d-M_{\ast}$ residual (shown as inset) is simply 
a scatter plot.  As noted by \cite{cou99} and \cite{cou03},
the lack of correlation between TF residual and disk scale 
length argues against the ``maximal disk'' hypothesis, 
in which the stellar disk provides a large fraction of 
the rotational support at $2.2R_d$, since in this case 
$V_{2.2}^2 \propto G M_{\ast} / (2.2 R_d)$.  We concur with both 
their observational result (indeed, our residual correlation 
appears even weaker) and with their conclusion.  

Filled squares in Figure 4 show model predictions 
for disks with mass fraction $m_d=0.05$ and 
the spin parameter $\lambda=0.06$ that yields the central 
solid line of Figure 3, with total halo masses of 
$M_h = 4\times 10^{10} M_\odot$, 1.89$\times 10^{11}M_\odot$, and 
1.39$\times 10^{12}$ $M_{\odot}$.  
(Following MMW, we define the halo mass within a virial radius
whose mean interior density is 200 times the critical density.)
We again assume an initial 
NFW halo concentration $c=10$ and compute $R_d$ using MMW's 
equation (28), but we compute the response of the halo to 
the disk using the improved adiabatic contraction approximation 
of \cite{gne04}, with code kindly provided by Oleg 
Gnedin.  With these parameters, the model reproduces 
the slope and normalization of our measured $M_{\ast}-V_{2.2}$ 
relation as well as the $R_d$-$M_{\ast}$ relation.  Horizontal 
lines attached to these points show the effect of changing the 
spin parameter to $\lambda=0.04$ and $\lambda=0.08$, corresponding 
to the lower and upper lines in Figure 3.  Larger disks have 
weaker self-gravity and therefore lower $V_{2.2}$, while compact 
disks make a substantial contribution to $V_{2.2}$ and therefore 
spin faster.  However, the predicted difference between large 
and small disks is small enough that it could plausibly be 
swamped by the scatter seen in Figure 4.  

Filled triangles in Figure 4 show models with $m_d=0.025$ and 
$\lambda=0.045$, which also reproduce the central line of 
Figure 3, for the same three halo masses (and $c=10$).  
Lowering $m_d$ reduces both $M_{\ast}$ and the disk contribution 
to $V_{2.2}$, but the shift is not exactly parallel to 
the $M_{\ast}-V_{2.2}$ relation, so the model predictions for 
$m_d=0.025$ lie below the central trend of the data, by 
$\sim0.1-0.2$ dex.  However, a choice of IMF with fewer low 
mass stars would reduce $M_{\ast}/L_i$ ratios at fixed 
color and could bring down the data points to agree with the $m_d=0.025$
predictions.   Because of the lower disk mass fraction, variations of
$\lambda$ that reproduce the spread in the $R_d-M_{\ast}$ 
relation ($\lambda=0.03-0.06$) produce only small shifts in $V_{2.2}$ 
at fixed $M_{\ast}$.  

Filled circles show models with $m_d=0.10$ and $\lambda=0.08$.  The 
model predictions now lie above the central trend of the data.  
Adopting a more bottom-heavy IMF could raise the data points 
and remove this discrepancy, but the $m_d=0.10$ model predicts 
a substantial change of $V_{2.2}$ over the range $\lambda=0.055-0.11$
that reproduces the spread in the $R_d$-$M_{\ast}$ relation.  
For $m_d=0.10$, the compact galaxies (open circles) should lie 
noticeably to the right of the large galaxies (open squares) in 
Figure 4, and they do not.  In other words, $m_d=0.10$ disks in 
adiabatically contracted, $c=10$, NFW halos are too close to maximal 
to be consistent with the absence of a size-TF residual correlation.  
It is difficult to put this discrepancy in fully quantitative 
terms because a viable model would have to specify what parameters 
{\it other} than $\lambda$ are varying to produce the intrinsic 
scatter in the $M_{\ast}-V_{2.2}$ relation. 
We will investigate this question in future work.

Figure 5a repackages the information in Figures 3 and 4 
by plotting the rotation velocity predicted for the stellar disk, 
\begin{equation}
V_{\ast,2.2} = \left({GM_* \over 2.2 R_d}\right)^{1/2}
  \left[1.32 \times 0.65 f_d + (1-f_d)\right]^{1/2}~,
\end{equation}
against the observed rotation velocity $V_{2.2}$.  
The factor 0.65 in the brackets is 
the fraction of the disk mass within $2.2R_d$, and the factor 
of 1.32 accounts for the flattened geometry of the disk potential 
(\citealt{fre70}; \citealt{bin87}).  The $(1-f_d)$ term
represents the contribution of the bulge, which 
we assume to lie entirely within $2.2R_d$; this contribution is small, 
since $f_d \ge 0.9$ for our sample.  In Figure 5a, we use the {\it mean} 
value of $R_d$ at the galaxy's $M_{\ast}$, from the best-fit relation 
shown in Figure 3 and listed in Table 1.  This plot simply tilts the 
stellar mass TF relation to account for the increase of average disk 
size with stellar mass.  The solid line shows the best-fit (forward) 
relation,
\begin{equation}
\label{eqn:vstar}
\frac{V_{\ast,2.2}}{156 \kms}=
  (0.64\pm0.01)\left(\frac{V_{2.2}}{156 \kms}\right)^{1.16\pm0.05}.
\end{equation}
The slope is just slightly steeper than the unit slope predicted for pure 
self-gravitating disks, but the typical offset is a factor 
of $\sim 0.65$.  Since velocities add in quadrature, 
$V_{\rm tot}^2=V_{\ast}^2+V_{h}^2$, 
the normalization of equation (9) implies a typical ratio of 
$V_{h}/V_{\ast}$ $\sim$ 1.2 of the halo and stellar disk 
circular velocities at $2.2R_d$.  The intrinsic scatter in $V_{\ast,2.2}$ 
at fixed $V_{2.2}$ is 0.057 dex.  If we apply \citeauthor{kan04}'s
(\citeyear{kan04}) color-based estimate of gas-to-stellar mass ratios
to compute the total disk contribution $V_{d,2.2}$ instead of
$V_{*,2.2}$, then the normalization of equation~(\ref{eqn:vstar})
rises slightly, to 0.71, and the slope changes to 1.07, a nearly
constant ratio of disk mass to dark halo mass within $2.2 R_d$.

In Figure 5b, we use each galaxy's actual scale length, 
instead of the mean scale length from the $R_d-M_{\ast}$ 
relation, when computing $V_{\ast,2.2}$.  The slope and 
normalization of the best-fit relation are virtually 
unchanged, but the intrinsic scatter is nearly two times larger,
0.105 dex instead of 0.058 dex.  Furthermore, the compact 
galaxies (circles) lie systematically above the mean relation 
(high $V_{\ast,2.2}$ at a given $V_{2.2}$), and the large 
galaxies (squares) lie systematically below.  Thus, even 
though the slope of the $V_{\ast,2.2}$-$V_{2.2}$ relation 
is close to unity, the value of $V_{\ast,2.2}$ is, on a galaxy-by-galaxy 
basis, a worse predictor of circular velocity than the 
stellar mass alone.  
Dashed lines in Figure 5 show $V_{*,2.2} = 0.85 V_{2.2}$, which
\cite{sac97} describes as a good approximation to the standard
``maximal disk'' hypothesis.\footnote{Specifically, \cite{sac97}
states that in maximal disk decompositions, the disk rotation speed
is typically 75-95\% of the rotation speed at $2.2 R_d$, with the
low end of the distribution populated by galaxies with large bulges,
which would be absent from our sample.}
Our direct estimates with the \cite{kro02} IMF lie below the 
maximal disk prediction, and they would continue to do so with the 
\cite{kan04} gas correction.  The increase
of scatter between Figures 5a and 5b,
another manifestation of the uncorrelated TF and disk size residuals,
implies that this gap is 
not simply a consequence of underestimating $M_*/L$ ratios;
dark halos must provide an important contribution to rotational
support at $2.2R_d$.

Figure 6a plots the inferred ratio of the total mass 
within 2.2 disk scale lengths to the stellar mass 
within 2.2 disk scale lengths, as a function of stellar 
mass.  The total mass is $M_{h,2.2}+M_{\ast,2.2}$ with
\begin{equation}
M_{h,2.2}=\frac{2.2 R_d}{G}(V_{2.2}^2-V_{\ast,2.2}^2),
\end{equation}
$M_{\ast,2.2} = 0.65 f_d M_{\ast} + (1-f_d) M_{\ast}$, and 
$V_{\ast,2.2}$ given by equation (8).  The total-to-stellar mass ratio 
has a flat trend with considerable 
scatter for $M_{\ast}$ $\ge$ $10^{10} M_{\odot}$, 
with a median value of $\sim$ 2.4.  For $M_{\ast}$ $<$ $10^{10} M_{\odot}$, 
the median ratio is higher ($\sim$ 4.4) in agreement with
previous results  (e.g., \citealt{per96}), and the scatter is larger.  
This increased mass ratio corresponds to the steeper than 
unit slope of the $V_{\ast,2.2}$-$V_{2.2}$ relation in Figure 5;
the increase is reduced but not eliminated if we use the
\cite{kan04} gas correction to 
estimate baryonic masses instead of stellar masses.
Points in Figure 6a are again coded by residual from the 
$R_d$-$M_{\ast}$ relation, and the separation of circles (compact galaxies) 
and squares (diffuse galaxies) shows that the larger galaxies, 
at fixed $M_{\ast}$, have higher halo-to-stellar mass ratios within 
$2.2 R_d$, as one might expect.  

While the scatter in disk sizes explains 
some of the scatter in $M_{h,2.2}/M_{\ast,2.2}$, 
Figure 6b shows that much of the scatter must arise 
from another source.  Here we plot
$M_{h,2.2}/M_{\ast,2.2}$
against $R_d/{\bar R_d (M_{\ast})}$, where
${\bar R_d (M_{\ast})}$ is the mean disk scale 
length at the galaxy's stellar mass based on 
the best-fit $R_d$-$M_{\ast}$ relation (dotted 
line of Figure 3).  Points are now coded by galaxy 
stellar mass, with $M_{\ast} < 10^{10} M_{\odot}$ 
shown by pentagons, $M_{\ast} > 10^{10.7} M_{\odot}$ 
shown by stars, and intermediate mass objects shown 
as crosses.  At a fixed $R_d/{\bar R_d (M_{\ast})}$, 
there is substantial scatter in
$M_{h,2.2}/M_{\ast,2.2}$.  
The systematically higher halo-to-stellar mass ratios 
of low mass galaxies are even clearer here than 
in Figure 6a.  

The three solid curves in Figure 6b show the
predictions of models with $m_d$=0.05 and NFW halo
concentrations of $c=5$, 10, and 20.  Along each
sequence, the value of $\lambda$ determines the value
of $R_d$, and we set ${\bar R_d (M_{\ast})}$ to be
the model prediction for $\lambda$=0.06, tracking
the central solid line of Figure 3.  For these calculations
we have assumed $M_{\ast}= 10^{10} M_{\odot}$, 
but the curves are the same for any choice of 
$M_{\ast}$.  To the extent that there is a central trend of
the data points, the model curves describe it reasonably well,
but it appears that a range of halo concentrations cannot account
for the large scatter in $M_{h,2.2}/M_{\ast,2.2}$ at fixed disk
size.  The upper and lower dotted curves show predictions
for $m_d$ = 0.025 and $m_d$=0.10, respectively, with $c$=10
and ${\bar R_d (M_{\ast})}$ computed assuming $\lambda$=0.045
(for $m_d$=0.025) and $\lambda$=0.08 (for $m_d$=0.10).  The envelope of
these curves contains most of the data points, though there
are a few with lower dark matter fractions.  In the context of
MMW-style disk galaxy models, where $m_d$, $c$, and $\lambda$ are the
parameters that determine disk properties, a substantial spread in
$m_d$ is required to explain the observed distribution of galaxies in the
($M_{\ast}$, $V_{2.2}$, $R_d$) space.  
In principle, the rotation curve shape can provide additional
constraints on the disk mass fraction and its variation with
galaxy properties (e.g., \citealt{persic88,per96}).
However, our rotation curves are not very well resolved spatially, 
and modeling rotation curve shapes requires specific assumptions about 
dark halo profiles, so we have not attempted to exploit these constraints 
here.

\cite{zav03} analyze the ratio of total mass to baryonic mass
in a literature sample of disk galaxies and conclude that it correlates
more directly with surface mass density than with luminosity
or scale length individually.  Figure 7 plots this
correlation for our data set, using the mean stellar surface density
within 2.2 scale lengths as the surface density measure.
There is indeed a steady correlation over nearly two orders of 
magnitude in surface density, somewhat tighter than the correlations
in Figure 6, though still with significant scatter
and occasional large outliers.  
Most significantly, one can see that the total-to-stellar mass ratios for 
low surface density, high mass galaxies are similar to those of typical
low mass galaxies with similar surface density.
This result suggests, as argued by \cite{zav03}, that the dependences
of the mass ratio on galaxy mass and scale length can be understood
as largely reflecting a more fundamental dependence on surface density.
The distribution of our data points in Figure 7 is similar to
 the distribution found by \cite{zav03}, though their
sample is constructed to include more low surface brightness galaxies.

Large filled squares show model predictions for $m_d=0.05$, 
halo masses 
$M_h = 4\times 10^{10} M_\odot$, 1.89$\times 10^{11}M_\odot$, and 
1.39$\times 10^{12}$ $M_{\odot}$,  
and spin parameter $\lambda = 0.06$.
Diagonal lines attached to these points span the range
$\lambda = 0.04-0.08$, with low-spin disks having high
$\Sigma_*$ and low $M_{h,2.2}/M_{*,2.2}$.
Filled triangles and circles show corresponding predictions
for $m_d=0.025$ and 0.05, with halo mass
1.89$\times 10^{11}M_\odot$ and the $\lambda$ values that
reproduce the three $R_d-M_*$ lines in Figure 3.
The data roughly follow the trend predicted for $\lambda$ variations
or $m_d$ variations as drivers of surface brightness variations.
At fixed $m_d$ and $\lambda$, a change of halo mass does not change
$M_{h,2.2}/M_{*,2.2}$, so the continuity of the trend for 
different mass galaxies again suggests that high mass galaxies
have higher $m_d$, not simply higher $M_h$.

\section{Summary and Discussion}
We have examined the correlations among stellar mass,
disk scale length, and rotation velocity at $2.2 R_d$ for
a sample of 81 disk-dominated galaxies selected from
the SDSS main galaxy redshift sample.   The SDSS
selection allows us to choose systems with a roughly
flat distribution of absolute magnitude over the
range $-18 \leq M_r \leq -22$ at redshifts such that
peculiar velocities induce relatively small distance uncertainties.
SDSS multi-color photometry allows
us to assign stellar masses to galaxies based on their
$i$-band luminosities and $g$-$r$ colors, using the prescription of
\cite{bel03} converted to a \cite{kro02}
IMF.   We use the SDSS $i$-band images to perform 2-d
bulge-disk decomposition with GALFIT (\citealt{pen02}).  
The defining morphological characteristics of our sample
are SDSS isophotal axis ratio $b/a$ $<$ 0.6 and GALFIT
$i$-band disk-to-total luminosity ratios $f_d\ge 0.9$. 
We do not apply any environmental pre-selection,
so our sample should be representative of the range
of environments in which galaxies of these
morphological characteristics and absolute magnitudes
are found.

Our principal observational results are as follows:
 
1.
We find a best-fit (forward) $i$-band TF relation
$(L_i/10^{10}L_\odot)=1.84 (V_{2.2}/150 \kms)^{2.60}$, with an
estimated
intrinsic scatter of 0.13 dex, or 0.33 magnitudes. 
The slope is shallower than that found by some previous
analyses (e.g., \citealt{ver01}; \citealt{kan02}) but
is similar to that of \cite{cou97}, whose sample selection
and analysis methods are closest to those here.
Possible sources of difference
include the environmental properties of the sample
(``field" vs. ``cluster"), 
the morphological criteria (disk/total $\geq 0.9$ vs. more general
``disk galaxy'' selection),
the adopted velocity measure
($V_{2.2}$ vs. $V_{\rm flat}$ or HI line width), and the fitting
procedures. 
The intrinsic scatter is similar to that found for
previous samples, though it rises substantially if we
do not restrict the sample to disk-dominated galaxies (see P05).
There is a weak trend for galaxies with redder than
average colors to lie below the best-fit TF relation (low $L_i$).

2.
We find a best-fit (forward) stellar mass TF relation
$(M_{\ast}/10^{10} M_{\ast}) = 2.32 ( V_{2.2}/ 156 \kms)^{3.05}$,
with an estimated intrinsic scatter of 0.16 dex.  
The use of stellar mass removes the
weak trend with color residual, suggesting that stellar mass plays
a more fundamental role than luminosity in TF correlations. 
The slope is shallower than that found by BD01 for Ursa Major spirals, 
with good agreement at $V_{2.2} \sim 200\kms$ but higher $M_*$
in our sample at $V_{2.2} \sim 100\kms$.  Possible sources of
difference again include range of environments, morphological
criteria, velocity measure, and fitting procedures.  
Our relation is also shallower than the {\it baryonic} mass 
TF relation of \cite{mcg00}, whose sample is more heterogeneous
and extends to lower luminosities.  
For our sample,
a statistical, color-based correction for gas-to-stellar mass
fractions \citep{kan04} makes only a modest difference to the
TF parameters.
 
3.
The $R_d-M_{\ast}$ distribution has a best-fit mean relation
$R_d= 4.0(M_{\ast}/2.2\times10^{10} M_{\odot})^{0.24}$ kpc, but
the distribution is broad, with roughly a factor of three range in
disk scale length at fixed $M_{\ast}$.  

4.
There is no discernible correlation between
residuals of the $R_d-M_{\ast}$ relation and residuals of
the $M_{\ast}-V_{2.2}$ relation. At a given $M_{\ast}$, compact
galaxies do not rotate faster or slower than average.  This
result agrees with earlier analyses showing weak or negligible
trends of TF residual with disk
scale length or surface brightness (\citealt{zwa95}; \citealt{cou99};
\citealt{ver01}). As emphasized by \cite{cou99}, the lack of
correlation
between TF residual and disk size implies that disks
cannot, in most cases, make a dominant contribution to rotation
velocities at $2.2 R_d$.  This evidence for ``sub-maximal" disks agrees
with independent arguments based on disk scale heights and
vertical velocity dispersions \citep{bot93,bot95,kre05}.
 
5.
Direct estimates of the stellar contribution
to the rotation speed at $2.2 R_d$, based on the population
synthesis mass-to-light ratios for a \cite{kro02} IMF,
yield a best-fit relation 
$(V_{\ast,2.2}/156 \kms) = 0.64(V_{2.2}/156\kms)^{1.16}$. 
Including estimated gas masses following \cite{kan04}
changes the intercept to 0.71 and the slope to 1.07.
The low contribution from the stellar disk is consistent with the weak
$R_d$-trend noted above, and it agrees well with 
\citeauthor{kre05}'s (\citeyear{kre05})
estimate of $0.53\pm 0.04$  based on disk scale heights. 
\cite{cou99} give an estimate of 0.6, also in good agreement 
with our results.
The scatter between $V_{\ast,2.2}$ and $V_{2.2}$ is larger
(0.11 dex vs. 0.06 dex) if we use each galaxy's actual $R_d$
instead of the mean $R_d$ at the galaxy's $M_{\ast}$, 
another sign of the weak correlation between size and rotation
speed at fixed $M_*$.

6.
The ratio of halo-to-stellar mass within $2.2 R_d$ has a large range
at a given $M_{\ast}$.  Median ratios are 2.4 for galaxies with
$M_{\ast} > 10^{10} M_{\odot}$ and 4.4 for $10^9 M_{\odot} <
M_{\ast} < 10^{10} M_{\odot}$. 
At any $M_{\ast}$, compact galaxies have lower
$M_{h,2.2}/M_{\ast,2.2}$
and large galaxies have higher $M_{h,2.2}/M_{\ast,2.2}$, as expected.
However, the spread in radius accounts for only a fraction of the scatter
in halo-to-stellar mass ratio. 
Galaxies with a wide range of $M_*$ and $R_d$ trace out a 
continuous correlation between halo-to-stellar mass and 
disk surface density, in agreement with \cite{zav03}, but
the scatter about the mean trend is substantial.

For early-type galaxies, the ``fundamental plane'' is close 
to a virial relation for the stellar 
component, $\sigma^2 \sim GM_{\ast}/R_{\rm eff}$ 
\citep{djo87,dre87,ber02},
and the scatter of the bivariate $L-\sigma$ 
relation is significantly larger than the scatter about
the fundamental plane.  The near-virial form 
is naturally explained if stars dominate the
central gravitational potential that determines 
the observed velocity dispersion, with only a modest contribution 
from dark matter (see \citealt{rus05} and references therein).
For disk galaxies, our results 
show the opposite:  $V_{2.2}$ is better correlated with $M_{\ast}$ 
(or $L$) than with $GM_{\ast}/R_d$, making the $M_{\ast}-V_{2.2}$
an essentially edge-on view of the disk galaxy fundamental plane.
The minimal effect of disk size on $V_{2.2}$ implies
that dark matter must contribute a large fraction of the
mass within the central two scale lengths of disk galaxies,
in accord with our direct (but IMF-dependent)
inference based on stellar mass-to-light ratios.
Our conclusions on these points agree with those of \cite{cou99}
and with the more recent analysis of \cite{cou03}, who investigate
the scaling relations of disk galaxy properties in a larger but
less tightly defined sample.

To put our observational results in context, we have compared
them to the predictions of theoretical models in which disks
form by the dissipative collapse of gas in cold dark matter
halos (\citealt{fal80}; \citealt{dal97};
MMW; \citealt{she02}; \citealt{dut05}). 
The scale length and rotation speed of a galaxy with
specified $M_{\ast}$ are determined in these models by the ratio
of disk mass to halo virial mass ($m_d$), the spin parameter
($\lambda$), and the concentration parameter ($c$) of the NFW halo. 
In practice, variations of $c$ within the expected range
have only moderate impact on $R_d$ and $V_{2.2}$, in part because adiabatic
contraction alters the inner regions of the dark
matter halo.  A model with $m_d=0.05$ reproduces the observed
$R_d-M_{\ast}$ distribution if $\lambda$ values range from
$\sim 0.04$ to $\sim 0.08$, a span that omits the lower half
and the extreme upper tail of the log-normal $\lambda$ distribution
predicted for dark matter halos (e.g. \citealt{bul01}).  
For $m_d=0.025$ and $m_d=0.10$, the required values of $\lambda$ 
are, respectively, lower by 25\% and higher by $\sim 30\%$.
 
The $m_d=0.05$ model reproduces our measured $M_{\ast}-V_{2.2}$
relation reasonably well, and it predicts a weak correlation
 between size and TF residual that could plausibly be washed
out by the TF scatter.  (Note, however, that such a model
with fixed $m_d$ does not explain the magnitude of the inferred
intrinsic scatter.)  The $m_d=0.025$ model predicts lower
$M_{\ast}$ at a given $V_{2.2}$, so it would require a different IMF
(with fewer low mass stars) to be consistent with our data. 
The $m_d=0.10$ model predicts slightly higher $M_{\ast}$ values at a
given $V_{2.2}$, and it predicts substantial size-TF residual
correlations because of its high mass disks.  These
strong residual correlations appear incompatible with our data. 
 
The model comparisons in Figure 6b suggest that no model
with a single value of $m_d$ will reproduce our observed distribution
of data points:  the large scatter in $M_{h,2.2}/M_{\ast,2.2}$
at fixed $R_d/{\bar R_d (M_{\ast})}$ can only be explained with
scatter in the ratio of stellar mass to {\it total} halo
mass. For our \cite{kro02} IMF normalization, Figure 6b
suggests $m_d$ values in the range $\sim 0.025-0.05$ for galaxies
with $10^9 M_{\odot} < M_{\ast} < 10^{10} M_{\odot}$, and
$\sim 0.05-0.10$ for galaxies with $M_{\ast} \ge 10^{10} M_{\odot}$.
Systematically lower $m_d$ values for low mass galaxies
are a plausible signature of more efficient
supernova feedback in shallow gravitational potential wells
\citep{dek86} or in galaxies that lack a confining envelope
of shock-heated gas \cite{ker04}.
 
We have not attempted to develop a complete and
self-consistent model for the distribution of disk
mass functions and halo parameters needed to reproduce
the observed, joint distribution of $M_{\ast}$, $R_d$,
and $V_{2.2}$.   The results above suggest the outline of
such a model:  disk mass fractions span a substantial range, with
a central value $m_d$ $\sim$ 0.05 and systematically lower
values for low mass galaxies, and the disk galaxy population
samples mainly the upper half of the halo spin parameter
distribution.  Our observations provide the data needed to
constrain more complete models of this sort, along the
lines pursued by \cite{she02} and \cite{dut05},
or to test the predictions of hydrodynamic simulations of
disk galaxy formation (e.g. \citealt{nav97a}). 
Confrontations between these models and our data should
lead to better understanding of the mechanisms that govern
disk galaxy formation and the relations between dark
and luminous matter in the inner regions of disk galaxies.

\acknowledgments

We thank Richard Pogge for assistance with the MDM observations and
Oleg Gnedin for helpful discussions and for providing his adiabatic
contraction code.  We also thank Michael Blanton and David 
Hogg for comments on the manuscript.
JP and DW acknowledge support from NSF Grants AST-0098584 and
AST-0407125.  

Funding for the creation and distribution of the SDSS Archive 
has been provided by the Alfred P. Sloan Foundation, the 
Participating Institutions, the National Aeronautics and 
Space Administration, the National Science Foundation, the 
U.S. Department of Energy, the Japanese Monbukagakusho, and 
the Max Planck Society. The SDSS Web site is http://www.sdss.org/.

The SDSS is managed by the Astrophysical Research Consortium 
(ARC) for the Participating Institutions. The Participating 
Institutions are The University of Chicago, Fermilab, the 
Institute for Advanced Study, the Japan Participation Group, 
The Johns Hopkins University, the Korean Scientist Group, Los 
Alamos National Laboratory, the Max-Planck-Institute for Astronomy
 (MPIA), the Max-Planck-Institute for Astrophysics (MPA), New Mexico 
State University, University of Pittsburgh, University of Portsmouth, 
Princeton University, the United States Naval Observatory, and
the University of Washington.

This paper is based in part on observations obtained in the framework
of the Calar Alto Key Project for SDSS Follow-up Observations (\citealt{gre01}) 
at the German-Spanish Astronomical Centre, Calar Alto Observatory, 
operated by the Max Planck Institute for Astronomy, Heidelberg, jointly 
with the Spanish National Commission for Astronomy.

\clearpage
\appendix

\section{Appendix}
Random errors in galaxy distances or magnitudes can lead to biased
estimates of TF parameters because there are more distant and faint
galaxies to scatter in one direction than vice versa, and because
objects can scatter across selection boundaries.  Most discussions
of these ``Malmquist''-type biases have focused on apparent
magnitude limited samples, or on the systematic bias in the
derived peculiar velocity field (e.g., 
\citealt{lyndenbell88,gould93,teerikorpi93,str95}).
Since our selection procedure is quite different from those of
most previous TF surveys, we have tested for Malmquist-type
biases with a simple Monte Carlo experiment.

The measurement errors in SDSS Petrosian magnitudes are generally
small, and luminosity uncertainties are therefore dominated by
line-of-sight peculiar velocities, which we have assumed in our
analysis to be drawn from a Gaussian of dispersion $300\kms$.
For our Monte Carlo sample, we assign random 3-d positions to $10^6$
artificial galaxies and draw their absolute $r$-band magnitudes from
the SDSS luminosity function of \cite{bla03c}.
We assign each of these galaxies a circular velocity drawn from
the best-fit $i$-band inverse TF relation listed in Table~2,
with slope $a=0.346$, intercept $b=\log V_{2.2} / \kms = 2.185$
at $\log L/L_\odot=10.293$, and intrinsic scatter of $\sigma = 0.048$ dex
in $\log V_{2.2}$.  For purposes of this experiment, we ignore the
slight difference between $i$-band and $r$-band TF relations, since
we only need a qualitatively realistic assignment for the test.

We modify each galaxy's redshift by a peculiar velocity drawn from
a $300\kms$ Gaussian, and we apply the same absolute-magnitude
dependent redshift cuts that we used for our sample definition
(see \S 2), using the {\it apparent} rather than true absolute magnitude.
Finally, we randomly draw from this cut sample a subset of galaxies
that matches the nearly flat $M_r$ distribution of our observed sample,
selecting ten times as many artificial galaxies as observed galaxies
in each absolute magnitude bin.
Applying our maximum likelihood estimation method to this artificial
sample yields inverse TF parameters $a=0.347$, $b=2.184$, and
$\sigma=0.050$, in excellent agreement with our input values.
The forward TF parameters are $a=2.613$, $b=10.26$, and $\sigma=0.14$ mag,
in excellent agreement with the values of
$2.603 \pm 0.133$, $10.266 \pm 0.020$, and $0.131 \pm 0.015$ derived
for our observed sample (see Table~2).
We conclude that any Malmquist-type biases in our sample definition
or analysis are smaller than our quoted statistical errors.
This small impact is not surprising, since the fractional distance
errors are small and the velocity and absolute magnitude range
of the sample are fairly large, making scatter across selection
boundaries a minimal effect.

\clearpage
\begin{deluxetable}{cccccccc}
\tabletypesize{\scriptsize}
\tablecaption{Galaxy Properties}
\tablewidth{0pt}
\tablehead{
\colhead{SDSS name} & 
\colhead{$d/{\rm Mpc}$ } &
\colhead{$L_{i}/10^{10}\,L_{\odot}$ } &
\colhead{$g$-$r$  } &
\colhead{$M_{\ast}/10^{10}\,M_{\odot}$} &
\colhead{$f_d$} &
\colhead{{$R_d/{\rm kpc}$} } &
\colhead{{$V_{2.2}/\kms$}}}
\startdata
J095743.26+004123.6 & 200.63 (4.43) &   4.86 (0.86) &0.49 (0.06) &  5.53 (1.07) & 0.98 &  4.62 (0.10) & 196.74 ( 2.87) \\ 
J100230.82+001826.2 & 145.15 (4.39) &   4.22 (0.43) &0.62 (0.03) &  6.13 (0.64) & 0.95 &  2.64 (0.08) & 266.68 ( 4.91) \\ 
J142729.65+010321.0 & 110.89 (4.37) &   0.56 (0.08) &0.32 (0.03) &  0.45 (0.06) & 0.96 &  2.42 (0.10) & 100.75 ( 3.87) \\ 
J144503.29+003137.1 & 124.82 (4.38) &   2.11 (0.32) &0.43 (0.04) &  2.12 (0.35) & 0.99 &  4.86 (0.17) & 157.89 ( 3.10) \\ 
J153045.16-002211.5 & 166.22 (4.41) &   4.57 (0.71) &0.54 (0.06) &  5.66 (0.95) & 0.93 &  3.69 (0.10) & 198.32 ( 2.58) \\ 
J232238.68-005903.7 & 148.36 (4.39) &   3.36 (0.35) &0.50 (0.02) &  3.88 (0.41) & 0.95 &  5.03 (0.15) & 168.21 ( 3.38) \\ 
J232613.88+010828.2 & 155.12 (4.40) &   2.68 (0.37) &0.41 (0.06) &  2.58 (0.42) & 0.99 &  3.81 (0.11) & 163.83 ( 3.22) \\ 
J233152.99-004934.4 & 106.31 (4.36) &   0.54 (0.10)\tablenotemark{a} &0.21 (0.08) &  0.35 (0.09) & 1.00 &  3.94 (0.17) &  82.23 ( 3.31) \\ 
J235603.89-000958.6 & 116.31 (4.37) &   1.02 (0.14) &0.25 (0.04) &  0.71 (0.11) & 0.97 &  3.58 (0.16) & 101.82 ( 2.67) \\ 
J234504.86-001615.1 & 103.61 (4.36) &   1.22 (0.23) &0.37 (0.06) &  1.09 (0.23) & 0.98 &  4.43 (0.19) & 116.33 ( 2.56) \\ 
J235624.68-001739.6 & 110.86 (4.37) &   0.41 (0.05) &0.39 (0.03) &  0.38 (0.05) & 1.00 &  2.02 (0.08) &  93.08 ( 8.01) \\ 
J001006.62-002609.6 & 143.75 (4.39) &   2.23 (0.31)\tablenotemark{a} &0.49 (0.03) &  2.54 (0.36) & 0.92 &  2.52 (0.08) & 130.02 ( 2.75) \\ 
J002025.78+004934.9 &  76.13 (4.34) &   2.06 (0.34) &0.74 (0.02) &  3.80 (0.63) & 0.92 &  5.85 (0.34) & 127.50 ( 7.68) \\ 
J004239.34+001638.7 & 195.48 (4.43) &   5.62 (0.59) &0.60 (0.03) &  7.86 (0.87) & 0.93 &  6.41 (0.15) & 246.59 ( 4.01) \\ 
J004935.68+010655.5 &  78.47 (4.34) &   4.05 (1.35) &0.56 (0.09) &  5.26 (1.83) & 1.00 &  7.95 (0.44) & 204.32 ( 2.01) \\ 
J094949.62+010533.2 & 151.02 (4.39) &   1.76 (0.26) &0.46 (0.05) &  1.86 (0.30) & 1.00 &  4.40 (0.13) & 145.83 ( 3.14) \\ 
J144307.79+010600.0 & 146.80 (4.39) &   2.03 (0.33) &0.43 (0.08) &  2.05 (0.40) & 0.95 &  3.21 (0.14) & 164.33 ( 3.32) \\ 
J015746.24-011229.9 & 187.72 (4.42) &   9.95 (2.25) &0.64 (0.06) & 15.01 (3.51) & 0.93 &  4.57 (0.11) & 323.99 (11.84) \\ 
J020853.01+004712.6 & 188.47 (4.42) &   7.96 (0.99) &0.57 (0.04) & 10.45 (1.38) & 0.98 &  6.05 (0.14) & 245.15 ( 3.87) \\ 
J022606.71-001954.9 &  94.38 (4.35) &   7.60 (1.23) &0.55 (0.05) &  9.59 (1.64) & 0.97 &  7.05 (0.33) & 251.28 ( 4.03) \\ 
J022751.44+003005.5 & 182.35 (4.42) &  10.03 (2.41) &0.49 (0.10) & 11.37 (3.06) & 0.93 &  9.88 (0.25) & 251.06 ( 2.78) \\ 
J022820.86+004114.0 & 183.32 (4.42) &   4.67 (0.43) &0.48 (0.03) &  5.19 (0.52) & 0.94 &  4.47 (0.11) & 220.11 ( 3.84) \\ 
J023610.91-005833.8 & 220.05 (4.44) &   8.70 (1.54) &0.57 (0.06) & 11.52 (2.17) & 0.94 &  5.92 (0.12) & 252.67 ( 6.44) \\ 
J211450.23-072743.3 & 127.16 (4.38) &   4.93 (0.51) &0.61 (0.02) &  7.05 (0.74) & 0.91 &  5.77 (0.21) & 217.62 ( 4.93) \\ 
J211439.91-075806.9 & 126.59 (4.38) &   1.51 (0.23) &0.39 (0.05) &  1.39 (0.23) & 1.00 &  3.78 (0.13) & 110.82 ( 3.81) \\ 
J211522.10-074605.0 & 124.79 (4.38) &   0.41 (0.04) &0.42 (0.04) &  0.41 (0.04) & 1.00 &  1.22 (0.04) &  55.98 ( 4.81) \\ 
J001708.77-005728.9 &  84.81 (4.35) &   0.81 (0.10) &0.48 (0.03) &  0.90 (0.12) & 1.00 &  3.14 (0.16) & 133.23 ( 5.56) \\ 
J024850.79-004602.6 &  95.50 (4.35) &   5.66 (0.96) &0.63 (0.05) &  8.38 (1.47) & 0.92 &  4.21 (0.19) & 200.45 ( 3.09) \\ 
J024728.01+003906.9 & 199.09 (4.43) &   4.97 (0.66) &0.65 (0.01) &  7.77 (1.03) & 0.92 &  3.35 (0.08) & 204.99 ( 3.95) \\ 
J213703.87-073518.0 & 131.33 (4.38) &   1.32 (0.15) &0.47 (0.03) &  1.44 (0.18) & 0.96 &  2.24 (0.08) & 135.39 ( 3.62) \\ 
J215421.67-075605.7 & 123.02 (4.37) &   0.92 (0.12) &0.47 (0.04) &  0.98 (0.14) & 0.98 &  1.42 (0.05) & 129.82 ( 2.66) \\ 
J112346.06-010559.4 &  75.88 (4.34) &   7.43 (1.06)\tablenotemark{a} &0.51 (0.03) &  8.74 (1.27) & 0.97 &  4.38 (0.25) & 264.51 ( 4.13) \\ 
J120155.64-010409.3 &  86.25 (4.35) &   2.74 (0.62) &0.69 (0.06) &  4.60 (1.06) & 0.97 &  3.54 (0.18) & 185.80 ( 3.83) \\ 
J124752.98-011109.0 &  99.32 (4.36) &   1.52 (0.35) &0.49 (0.07) &  1.72 (0.43) & 0.93 &  4.97 (0.22) & 141.43 ( 1.93) \\ 
J232021.17-001819.2 & 113.54 (4.37) &   0.44 (0.05) &0.36 (0.02) &  0.38 (0.04) & 1.00 &  3.09 (0.12) & 120.31 (73.27) \\ 
J233259.33+004318.8 &  78.69 (4.34) &   0.26 (0.04) &0.37 (0.05) &  0.23 (0.04) & 1.00 &  2.26 (0.13) &  80.12 ( 8.25) \\ 
J235106.25+010324.0 & 120.59 (4.37) &   7.95 (1.47) &0.61 (0.05) & 11.41 (2.20) & 0.97 &  8.80 (0.32) & 218.10 ( 5.29) \\ 
J020447.19+005006.3 &  88.98 (4.35) &   0.31 (0.05) &0.27 (0.06) &  0.22 (0.04) & 1.00 &  2.13 (0.11) &  78.20 ( 3.73) \\ 
J032019.21+003005.4 & 103.19 (4.36) &   0.19 (0.02) &0.30 (0.03) &  0.14 (0.02) & 1.00 &  1.72 (0.09) &  85.25 (10.45) \\ 
J080658.75+463346.8 &  97.38 (4.36) &   2.33 (0.52) &0.65 (0.06) &  3.62 (0.83) & 0.96 &  3.86 (0.17) & 165.67 ( 3.43) \\ 
J082956.27+515824.1 &  77.15 (4.34) &   0.40 (0.06) &0.26 (0.04) &  0.28 (0.05) & 1.00 &  3.23 (0.18) &  83.72 ( 3.39) \\ 
J033329.46-073308.4 &  76.72 (4.34) &   0.40 (0.06) &0.35 (0.03) &  0.34 (0.06) & 1.00 &  2.19 (0.12) &  97.42 ( 3.83) \\ 
J085705.72+514850.7 &  76.09 (4.34) &   0.35 (0.06) &0.30 (0.04) &  0.27 (0.05) & 0.92 &  5.08 (0.32) & 107.86 ( 4.80) \\ 
J211816.06-073507.8 & 128.66 (4.38) &   7.25 (1.20) &0.69 (0.05) & 12.05 (2.07) & 0.90 &  5.00 (0.17) & 228.53 ( 5.76) \\ 
J020045.13-101451.3 &  81.75 (4.34) &   0.18 (0.02) &0.37 (0.02) &  0.16 (0.02) & 1.00 &  2.40 (0.13) &  77.24 ( 8.71) \\ 
J082949.87+484647.9 & 105.70 (4.36) &   3.39 (0.71) &0.57 (0.07) &  4.53 (1.00) & 1.00 &  4.08 (0.17) & 141.97 ( 2.14) \\ 
J084408.09+504422.9 &  76.21 (4.34) &   0.69 (0.10) &0.51 (0.03) &  0.80 (0.12) & 0.99 &  2.66 (0.15) &  97.89 ( 5.30) \\ 
J125715.15-003927.5 &  98.28 (4.36) &   1.13 (0.17) &0.56 (0.06) &  1.48 (0.25) & 0.95 &  1.60 (0.07) & 138.34 ( 3.25) \\ 
J135433.67-004635.0 & 114.58 (4.37) &   1.45 (0.19) &0.62 (0.01) &  2.11 (0.27) & 0.95 &  2.12 (0.08) & 153.15 ( 3.85) \\ 
J133839.73+003245.0 &  97.29 (4.36) &   0.30 (0.04) &0.35 (0.03) &  0.26 (0.03) & 0.92 &  4.40 (0.21) & 106.86 ( 3.83) \\ 
J150546.86-004253.6 & 155.02 (4.40) &   2.88 (0.40) &0.55 (0.02) &  3.69 (0.52) & 0.99 &  4.93 (0.14) & 174.65 ( 2.50) \\ 
J143842.98-000027.9 & 146.25 (4.39) &   3.49 (0.67) &0.66 (0.04) &  5.53 (1.08) & 0.98 &  3.13 (0.09) & 200.67 ( 5.22) \\ 
J145025.02-011026.5 & 187.50 (4.42) &   7.23 (1.40) &0.50 (0.08) &  8.28 (1.81) & 0.96 &  5.70 (0.14) & 244.95 ( 3.27) \\ 
J140452.62-003640.5 & 106.16 (4.36) &   3.40 (0.36) &0.49 (0.02) &  3.79 (0.41) & 0.96 &  5.61 (0.23) & 185.34 ( 3.80) \\ 
J141413.17-005339.8 & 165.01 (4.40) &   8.11 (1.08) &0.54 (0.06) & 10.08 (1.53) & 0.91 &  6.80 (0.18) & 270.77 ( 3.22) \\ 
J232631.10+005013.5 & 127.60 (4.38) &   2.13 (0.28) &0.51 (0.03) &  2.49 (0.34) & 0.95 &  2.56 (0.09) & 162.55 ( 2.81) \\ 
J234328.26-000148.6 & 165.10 (4.40) &   2.62 (0.45) &0.63 (0.05) &  3.89 (0.68) & 0.95 &  3.71 (0.10) & 194.10 ( 3.52) \\ 
J024459.89+010318.5 & 112.33 (4.37) &   2.62 (0.51) &0.55 (0.05) &  3.31 (0.68) & 0.99 &  3.40 (0.13) & 185.06 ( 3.20) \\ 
J021859.65+001948.0 & 132.52 (4.38) &   3.04 (0.49) &0.59 (0.04) &  4.22 (0.69) & 0.98 &  3.20 (0.11) & 197.01 ( 3.68) \\ 
J025627.12+005232.6 & 101.39 (4.36) &   3.37 (0.39) &0.44 (0.05) &  3.46 (0.48) & 0.99 &  5.95 (0.26) & 160.82 ( 3.86) \\ 
J203523.80-061437.9 &  85.87 (4.35) &   4.32 (0.64)\tablenotemark{a} &0.59 (0.04) &  5.92 (0.91) & 0.97 & 10.76 (0.55) & 254.62 ( 2.95) \\ 
J204256.27-065126.1 & 126.41 (4.38) &   1.04 (0.16) &0.56 (0.05) &  1.34 (0.22) & 0.95 &  2.53 (0.09) & 117.70 ( 5.22) \\ 
J205532.62+000635.6 & 133.69 (4.38) &   0.74 (0.12) &0.15 (0.07) &  0.43 (0.10) & 1.00 &  2.88 (0.10) &  98.59 ( 2.91) \\ 
J205103.70+000825.5 & 102.06 (4.36) &   0.62 (0.09) &0.26 (0.04) &  0.44 (0.07) & 1.00 &  2.67 (0.13) & 112.83 (10.20) \\ 
J215156.74+121411.3 & 123.27 (4.37) &   1.02 (0.20) &0.48 (0.05) &  1.12 (0.24) & 0.94 &  2.79 (0.10) & 119.70 ( 5.22) \\ 
J205404.34+004638.6 & 130.08 (4.38) &   5.65 (0.96) &0.53 (0.06) &  6.97 (1.27) & 0.98 &  6.01 (0.21) & 217.05 ( 3.60) \\ 
J215652.70+121857.5 & 135.91 (4.38) &   3.84 (0.44) &0.54 (0.04) &  4.82 (0.58) & 0.92 &  5.68 (0.19) & 184.02 ( 3.71) \\ 
J211343.93+003428.7 & 212.02 (4.44) &  11.12 (2.46) &0.57 (0.06) & 14.75 (3.40) & 0.97 &  8.07 (0.25) & 283.32 ( 3.64) \\ 
J015946.76-001657.7 & 194.89 (4.43) &   5.45 (0.65) &0.58 (0.04) &  7.30 (0.91) & 0.95 &  4.93 (0.11) & 241.10 ( 3.92) \\ 
J012223.78-005230.7 & 120.89 (4.37) &  16.10 (2.64) &0.67 (0.06) & 25.81 (4.45) & 0.90 & 12.06 (0.44) & 297.68 ( 3.57) \\ 
J115731.83-011510.5 &  76.70 (4.34) &   1.47 (0.32) &0.45 (0.07) &  1.54 (0.37) & 0.98 &  3.95 (0.22) & 129.62 ( 2.57) \\ 
J205307.50-002407.0 & 132.87 (4.38) &   0.82 (0.09) &0.39 (0.03) &  0.76 (0.09) & 0.97 &  3.74 (0.17) & 114.02 ( 4.56) \\ 
J215247.62+122942.8 & 129.32 (4.38) &   0.88 (0.09) &0.64 (0.02) &  1.34 (0.13) & 0.93 &  3.32 (0.12) & 124.53 (20.89) \\ 
J215326.90+002218.0 & 123.77 (4.37) &   1.05 (0.12) &0.36 (0.04) &  0.93 (0.12) & 0.99 &  5.18 (0.19) & 135.87 ( 3.83) \\ 
J210633.54+104504.1 & 126.71 (4.38) &   1.21 (0.14) &0.48 (0.03) &  1.35 (0.17) & 1.00 &  3.51 (0.12) & 108.10 ( 4.02) \\ 
J210039.64-001236.6 & 107.64 (4.36) &   0.66 (0.12) &0.46 (0.06) &  0.70 (0.14) & 0.99 &  3.25 (0.16) & 102.23 ( 3.00) \\ 
J213811.68+121139.1 &  99.66 (4.36) &   1.47 (0.22) &0.46 (0.05) &  1.56 (0.26) & 0.91 &  2.04 (0.09) &  99.20 ( 3.43) \\ 
J124545.20+535702.0 & 102.17 (4.36) &   0.36 (0.05) &0.26 (0.07) &  0.25 (0.05) & 1.00 &  2.78 (0.12) &  89.83 ( 6.49) \\ 
J080046.85+353146.0 &  76.40 (4.34) &   0.26 (0.04) &0.37 (0.02) &  0.23 (0.03) & 1.00 &  1.61 (0.09) &  90.90 (13.37) \\ 
J120430.82+022036.1 &  84.17 (4.35) &   0.26 (0.03) &0.39 (0.02) &  0.24 (0.03) & 1.00 &  2.74 (0.15) &  85.49 (17.28) \\ 
J123058.64+513636.2 &  79.80 (4.34) &   0.38 (0.06) &0.43 (0.04) &  0.38 (0.06) & 1.00 &  2.53 (0.14) &  85.65 ( 7.77) \\
\enddata
\tablecomments{ {Local} Group barycenter distances are in Mpc, for
$H_0 = 70\hubunits$.
Luminosities are derived from SDSS Petrosian magnitudes and corrected
for internal dust extinction following \cite{tul98}.
Colors are SDSS model colors corrected for internal extinction.
The observed Petrosian magnitudes, model colors, 
helio-centric redshifts, and corrected frames
can be obtained from the public SDSS DR2 server at http://www.sdss.org/dr2.
}

\tablenotetext{a}{GALFIT magnitude used instead of SDSS Petrosian, 
because of poor pipeline estimation of the total observed magnitude.}
\end{deluxetable}

\clearpage
\begin{deluxetable}{rrrrrr}
\tabletypesize{\scriptsize}
\tablecaption{ Bivariate Relation Fits\label{tbl-1}}
\tablewidth{0pt}
\tablehead{
\colhead{$y$} & 
\colhead{$x$} &
 \colhead{$x_0$} &
\colhead{a($\pm$)} & 
\colhead{b($\pm$)} & 
\colhead{$\sigma$($\pm$)} }
\startdata
${\rm log} L_i$ & ${\rm log}  V_{2.2}$ & 2.175 & 2.603(0.133) & 10.266(0.020) & 0.131(0.015) \\
${\rm log}  V_{2.2}$ & ${\rm log}  L_i$ & 10.293 & 0.342(0.016) & 2.185(0.006) & 0.048(0.005) \\
$g$-$r$ & ${\rm log}  L_i$ & 10.293 & 0.179(0.013) & 0.479(0.009) & 0.073(0.010) \\
$g$-$r$ & ${\rm log}  M_{\ast}$ & 10.345 & 0.171(0.012) & 0.499(0.008) & 0.061(0.008) \\
${\rm log}   M_{\ast}$ & ${\rm log}  V_{2.2}$ & 2.192 & 3.048(0.121) & 10.365(0.018) & 0.158(0.021) \\
${\rm log}  V_{2.2}$ & ${\rm log}   M_{\ast}$ & 10.345 & 0.291(0.011) & 2.186(0.006) & 0.049(0.007) \\
${\rm log}   R_d$ & ${\rm log}   M_{\ast}$ & 10.345 & 0.242(0.030) & 0.588(0.012) & 0.142(0.011) \\
${\rm log}  V_{\ast,2.2}$\tablenotemark{a} & ${\rm log}  V_{2.2}$ & 2.192 & 1.157(0.053) & 2.000(0.008) & 0.057(0.008) \\
${\rm log}  V_{2.2}$ & ${\rm log}  V_{\ast,2.2}$\tablenotemark{a} & 1.93 & 0.765(0.042) & 2.138(0.007) & 0.047(0.007) \\
${\rm log}  V_{\ast,2.2}$\tablenotemark{b} & ${\rm log}  V_{2.2}$ & 2.192 & 1.128(0.085) & 2.000(0.011) & 0.105(0.014) \\
${\rm log}  V_{2.2}$ & ${\rm log}  V_{\ast,2.2}$\tablenotemark{b} & 1.93 & 0.661(0.050) & 2.143(0.010) & 0.081(0.011) \\
\enddata
\tablecomments{Bivariate relations are fit with 
the model $y$=$a$($x$-$x_0$)+$b$ with a Gaussian intrinsic 
scatter of $y$ at fixed $x$, with dispersion $\sigma$.  Errors
listed for $a$, $b$, and $\sigma$ are computed from 100 
bootstrap trials, and the value of $x_0$ is chosen so that 
errors in $a$ and $b$ are uncorrelated.  Luminosities are in 
$L_{i,\odot} $, stellar masses in $M_{\odot}$, velocities 
in ${\rm km\,s^{-1}}$, and radii in ${\rm kpc}$.  }

\tablenotetext{a}{$V_{\ast,2.2}$ is computed using the mean disk scale 
length for the galaxy's stellar mass.}
\tablenotetext{b}{$V_{\ast,2.2}$ is computed using the each galaxy's 
measured scale length.}
\end{deluxetable}

\clearpage
\begin{figure}
\epsscale{0.99}
\plotone{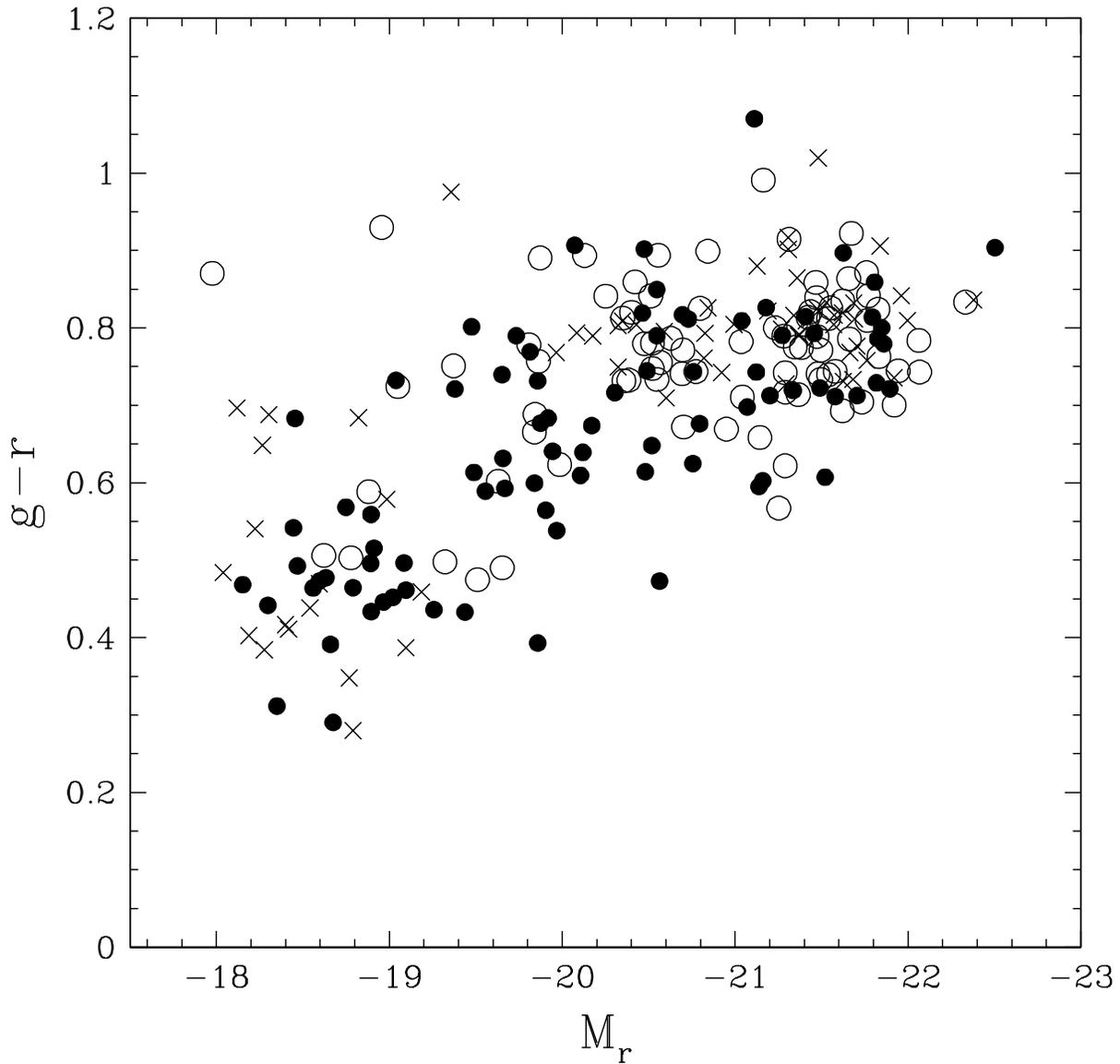}
\caption{Distribution of sample galaxies in the color-absolute magnitude plane.
Filled circles are the sample of 81 disk-dominated galaxies analyzed in this 
paper, and open circles are galaxies with $f_d$ less 
than 0.9.  Crosses show galaxies with insufficient extended H$\alpha$ 
for useful rotation curve measurements.
}
\end{figure}

\clearpage
\begin{figure}
\epsscale{0.80}
\plotone{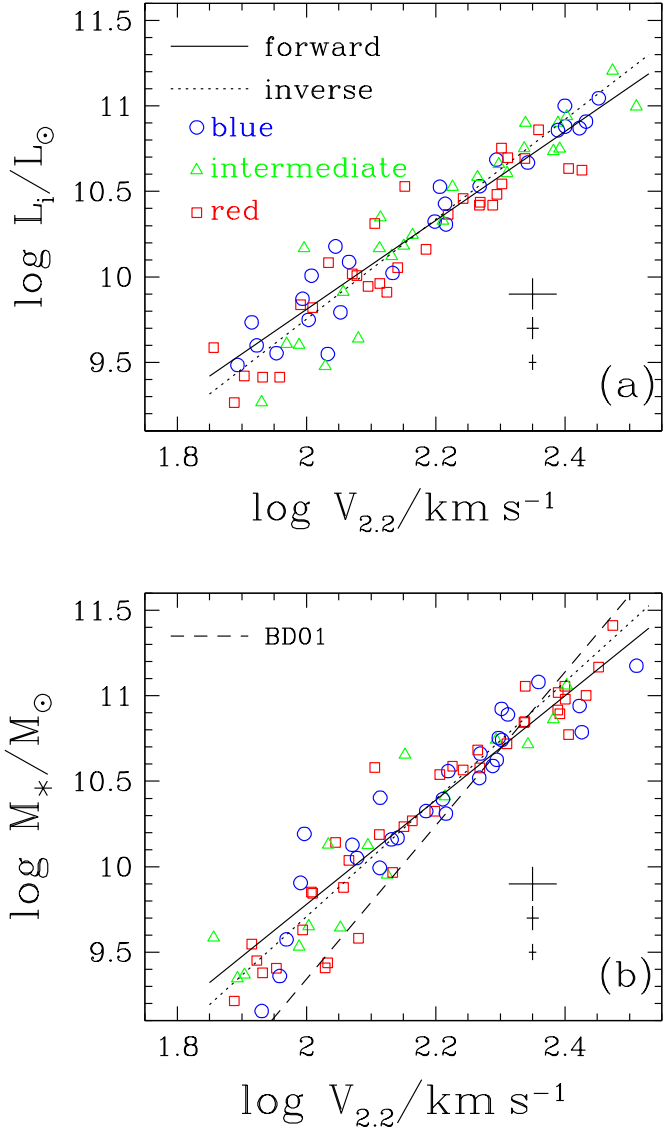}
\caption{
{\it (a)} TF relation of our disk-dominated sample: inclination corrected
$i$-band luminosity vs.\ circular velocity at 2.2 disk scale lengths.
Circles, triangles, and squares show galaxies that are blue, intermediate,
and red with respect to the mean color-absolute magnitude relation.
Solid and dotted lines show the best forward and inverse fits to the 
data points (see Table~2 for parameters).
{\it (b)} Same as {\it (a)}, but with $L_i$ replaced by the estimated stellar
mass $M_*$.  Here and in subsequent figures, $M_*$ is 
estimated from the $i$-band luminosity and a stellar 
mass-to-light ratio inferred from the $g$-$r$ color following 
\cite{bel03} (see equation 6).  The short-dashed line shows the best-fit 
stellar mass TF relation from BD01, adjusted to the Kroupa (IMF).
In this and all subsequent figures,
error crosses show the 90th, 50th, and 10th-percentile values of
the observational errors.
}
\end{figure}
\clearpage

\begin{figure}
\epsscale{0.9}
\plotone{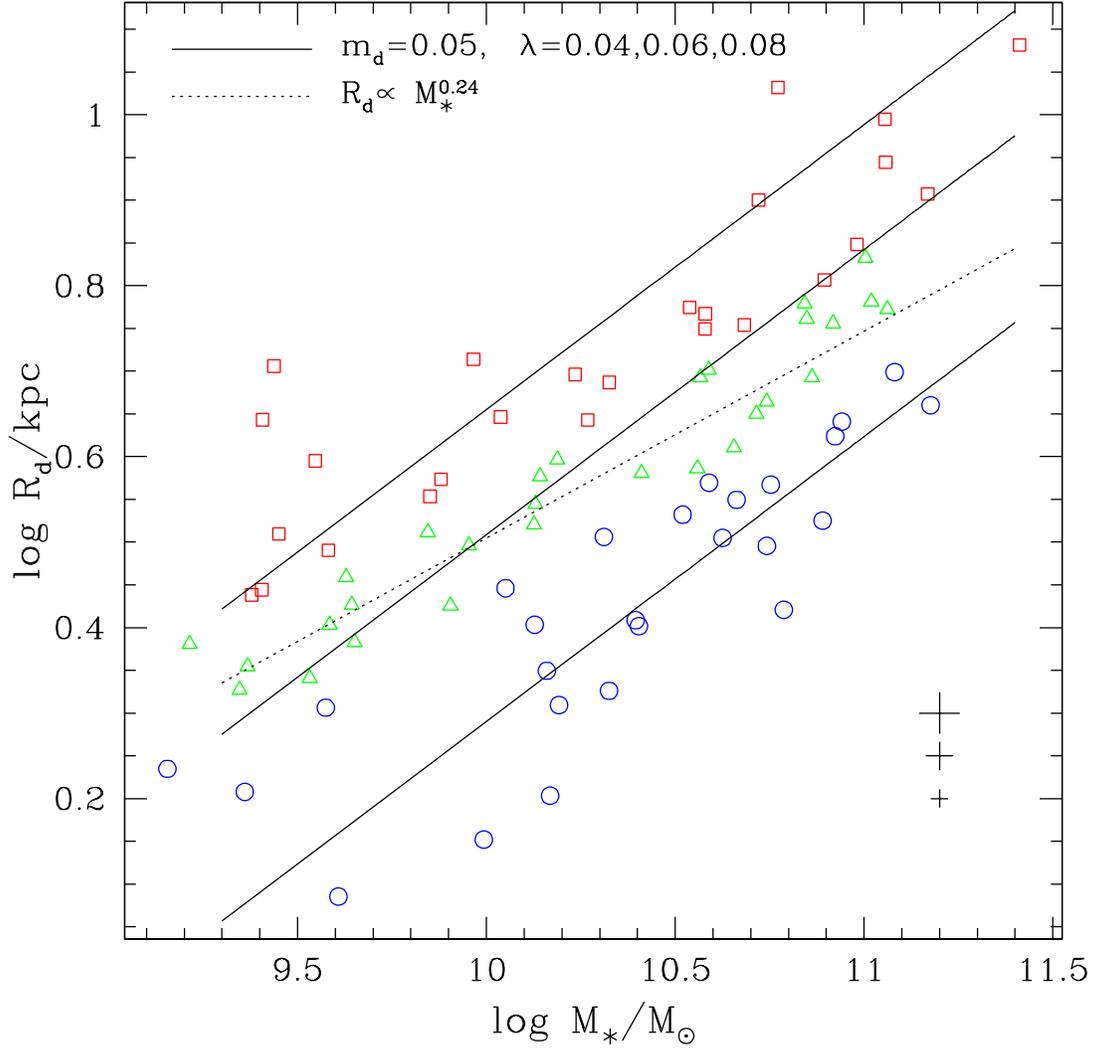}
\caption{ 
Relation between disk scale length and stellar mass.  The dotted line is
the best-fit mean relation, and points are coded by distance from this
relation.  The central solid line shows the predicted relation for 
exponential disks formed in adiabatically contracted NFW halos
with concentration $c=10$, spin parameter $\lambda=0.06$, and ratio 
$m_d=0.05$ of disk mass to halo virial mass.
Lower and upper solid lines show the model predictions for 
$\lambda=0.04$ and 0.08, respectively.
Other combinations of $(m_d,\lambda)$ can produce similar results,
as discussed in the text.
}
\end{figure}

\clearpage
\begin{figure}
\epsscale{1.0}
\plotone{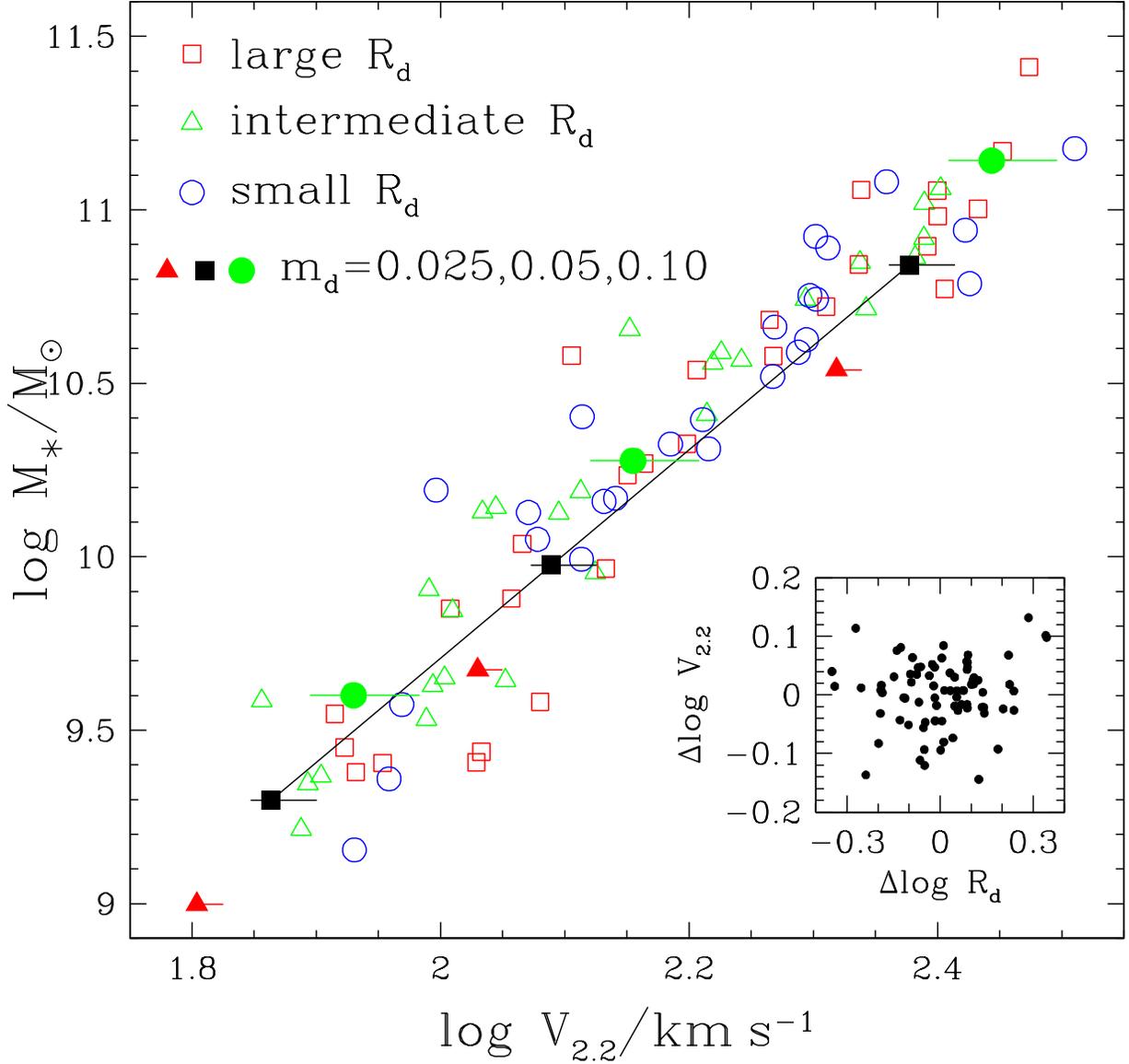}
\caption{
Stellar mass TF relation, as in Fig.~2b, but with points coded by
their residual from the mean $R_d-M_*$ relation, as in Fig.~3.
The inset panel plots TF residual against $R_d$ residual.
In the main panel, filled squares show predictions for a model with
$m_d=0.05$, $c=10$, and $\lambda=0.06$, for three different halo masses.
Horizontal error bars on these points show the change in $V_{2.2}$ 
when $\lambda$ is changed to 0.04 (higher $V_{2.2}$) or 0.08 
(lower $V_{2.2}$), the values corresponding roughly to the range of
observed disk sizes.  Filled circles and filled triangles show 
corresponding predictions for $m_d=0.10$ and $0.025$, respectively,
with $\lambda$ values chosen to produce the observed range of disk
sizes for the corresponding $m_d$ (see text for further discussion).
}
\end{figure}

\clearpage
\begin{figure}
\epsscale{1.0}
\plotone{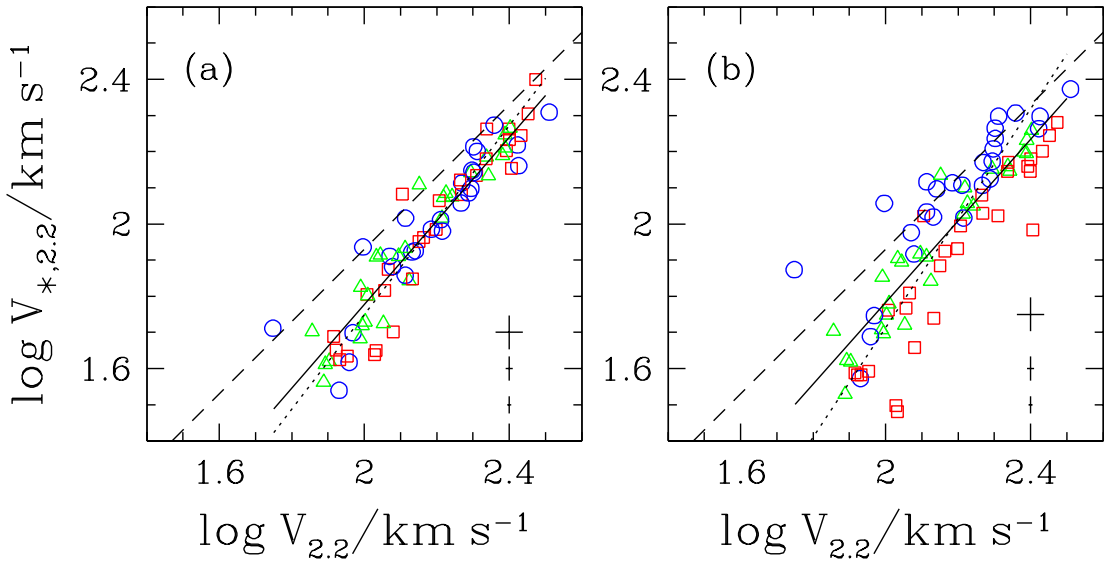}
\caption{
Comparison of the rotation velocity $V_{*,2.2}$ predicted for each
galaxy's stellar component (eq.~[8]) to the observed $V_{2.2}$.
Points are coded by residual from the $R_d-M_*$ relation, 
as in Fig.~3.  In panel {\it (a)}, $V_{*,2.2}$ is computed using the
{\it mean} $R_d$ at each galaxy's $M_*$, while in panel {\it (b)} it is
computed using each galaxy's observed $R_d$.  Solid and dotted lines
show forward and inverse fits with parameters listed in Table~2.
The dashed line shows the relation $V_{*,2.2}=0.85V_{2.2}$ expected
for ``maximal'' disks \citep{sac97}.
}
\end{figure} 

\clearpage
\begin{figure}
\epsscale{1.0}
\plotone{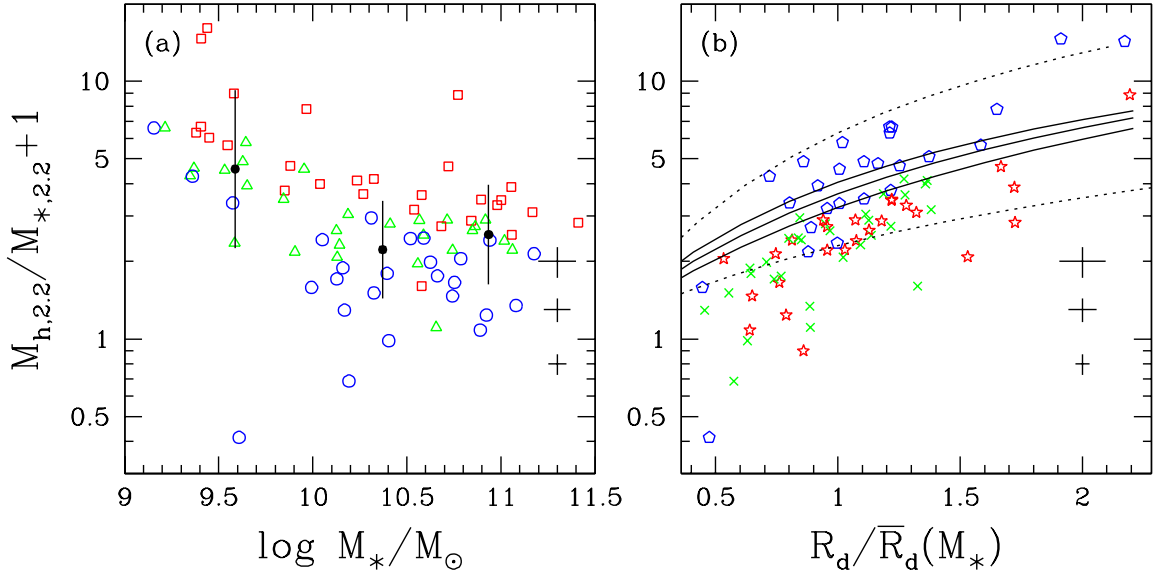}
\caption{ 
{\it (a)} Ratio of total mass within $2.2R_d$ to stellar mass within $2.2 R_d$,
as a function of $M_*$.
Points are coded by residual from the
$R_d-M_*$ relation.  Filled circles with error bars show the mean
and standard deviation for galaxies in the mass ranges
${\rm log}\, M_*/M_\odot > 10.7$, $10-10.7$, and $<10$.
{\it (b)} Similar to {\it (a)}, but mass ratios are now plotted against the
ratio of each galaxy's scale length to the mean scale length
for its stellar mass.  Crosses, stars, and pentagons show galaxies
with $\log\, M_*/M_\odot > 10.7$, $10-10.7$, and $<10$, respectively.
The three solid curves show predictions of models with 
$m_d=0.05$ and concentration parameters $c=5$, 10, and 20 (bottom
to top).  The dotted curves show predictions for $m_d=0.025$ (upper)
and $m_d=0.1$ (lower), for $c=10$.
 }
\end{figure}
\clearpage
\begin{figure}
\epsscale{0.999}
\plotone{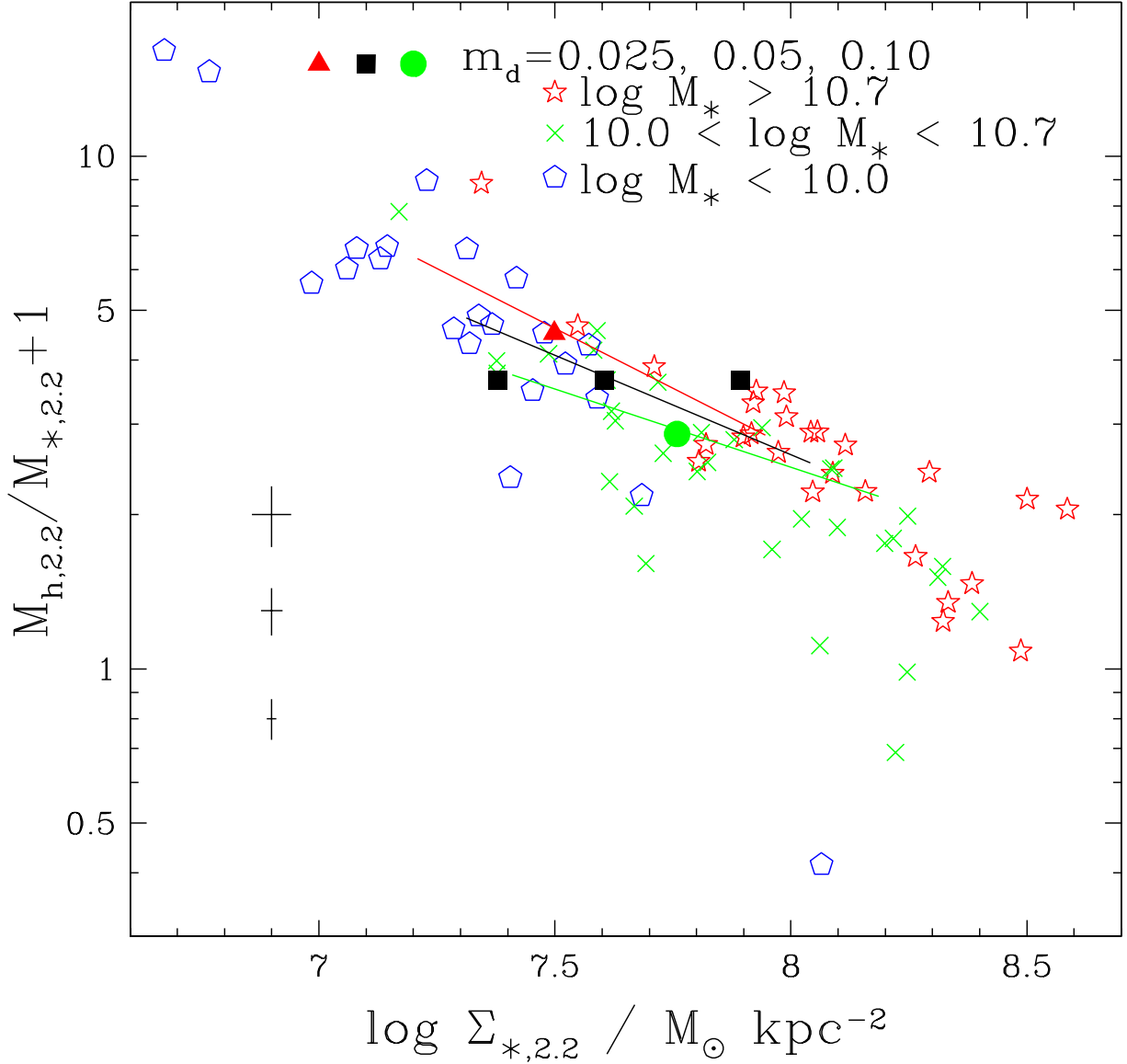}
\caption{
Total-to-stellar mass ratio as a function of the mean stellar surface
density within $2.2R_d$.  Stars, crosses, and pentagons represent
high, intermediate, and low mass galaxies, as in Fig.~6b.
The filled triangle, square, and circle with attached diagonal
lines represent models with $(m_d,\lambda)=(0.025,0.045)$, 
(0.05,0.06), and (0.10,0.08), respectively, with $c=10$
and a halo mass $M_h=1.89 \times 10^{11} M_\odot$.
The points correspond to the three central models shown in Fig.~4, 
and lines show the effect
of varying $\lambda$ over the range that reproduces the observed
range of $R_d$, as in Fig.~4.
Additional filled squares show
the $m_d=0.05$ model for $M_h=4\times 10^{10}M_\odot$ (low $\Sigma_*$)
and $1.39 \times 10^{12} M_\odot$ (high $\Sigma_*$); the lines
of varying $\lambda$ would approximately parallel those of
the central model.
}
\end{figure}

\end{document}